\documentclass[letterpaper,twoside,twocolumn,english,aps,pre,reprint]{revtex4-1}
\usepackage[T1]{fontenc}
\usepackage[latin9]{inputenc}
\usepackage{color}
\usepackage{array}
\usepackage{units}
\usepackage{amsmath}
\usepackage{graphicx}
\usepackage{amssymb}

\makeatletter

\pdfpageheight\paperheight
\pdfpagewidth\paperwidth

\providecommand{\tabularnewline}{\\}

 \@ifundefined{textcolor}{}
 {%
  \definecolor{BLACK}{gray}{0}
  \definecolor{WHITE}{gray}{1}
  \definecolor{RED}{rgb}{1,0,0}
  \definecolor{GREEN}{rgb}{0,1,0}
  \definecolor{BLUE}{rgb}{0,0,1}
  \definecolor{CYAN}{cmyk}{1,0,0,0}
  \definecolor{MAGENTA}{cmyk}{0,1,0,0}
  \definecolor{YELLOW}{cmyk}{0,0,1,0}
  }

\newcommand{\C}[1]{{}}

\makeatother

\usepackage{babel}

\begin{document}
\global\long\def\Tc{T_{\mathrm{c}}}
\global\long\def\Lp{L_{\parallel}}
\global\long\def\Ls{L_{\perp}}
\global\long\def\fex{f_{\mathrm{ex}}}
\global\long\def\Fc{\beta\mathcal{F}_{\mathrm{C}}}
\global\long\def\Fi{\beta\mathcal{F}_{\mathrm{I}}}
\global\long\def\xip{\xi_{+}}
\global\long\def\Xc{\vartheta}
\global\long\def\Dc{\Delta}
\global\long\def\Dcs{\Delta_{\perp}}
\global\long\def\Dcp{\Delta_{\parallel}}
\global\long\def\B{b}

\title{Aspect-ratio dependence of thermodynamic Casimir forces}

\author{Alfred Hucht, Daniel Grüneberg and Felix M. Schmidt}

\affiliation{Fakultät für Physik, Universität Duisburg--Essen, D--47048 Duisburg}
\begin{abstract}
We consider the three-dimensional Ising model in a $L_{\perp}\times L_{\parallel}\times L_{\parallel}$
cuboid geometry with finite aspect ratio $\rho=L_{\perp}/L_{\parallel}$
and periodic boundary conditions along all directions. For this model
the finite-size scaling functions of the excess free energy and thermodynamic
Casimir force are evaluated numerically by means of Monte Carlo simulations.
The Monte Carlo results compare well with recent field theoretical
results for the Ising universality class at temperatures above and
slightly below the bulk critical temperature $T_{\mathrm{c}}$. Furthermore,
the excess free energy and Casimir force scaling functions of the
two-dimensional Ising model are calculated exactly for arbitrary $\rho$
and compared to the three-dimensional case. We give a general argument
that the Casimir force vanishes at the critical point for $\rho=1$
and becomes repulsive in periodic systems for $\rho>1$.
\end{abstract}


\maketitle

\section{Introduction\label{sec:Introduction}}

The spatial confinement of a fluctuating and highly correlated medium
may cause long-range forces. The Casimir effect, which was theoretically
predicted in 1948 by the Dutch physicist H.\,B.\,G.\,Casimir \cite{Casimir48},
is a prominent example. This quantum effect is caused by the vacuum
fluctuations of the electromagnetic field and has been proven experimentally
in the late 1990's \cite{Lamoreaux97,MohideenRoy98}. It becomes manifest
in an attractive long-range \emph{Casimir force} acting between two
parallel, perfectly conducting plates in an electromagnetic vacuum.

Another example for a fluctuation-induced force being of similar nature
as the Casimir force in quantum electrodynamics can be found in the
physics of critical phenomena \cite{FisherdeGennes78,Gambassi09a}.
This \emph{thermodynamic Casimir effect} is caused by the spatial
confinement of thermal fluctuations near the critical point of a second
order phase transition. Experimentally, it has been proven for the
first time by measuring the film thickness of superfluid $^{4}$He
films as a function of the temperature in the vicinity of the lambda
transition \cite{GarciaChan99,GanshinScheidemantelGarciaChan06}.
Since then, the thermodynamic Casimir effect was measured in several
different systems including binary liquid mixtures \cite{FukutoYanoPershan05,HertleinHeldenGambassiDietrichBechinger08,Gambassi09}
and tricritical $^{3}$He-$^{4}$He \cite{GarciaChan02}.

For several years the shape of the finite-size scaling function determined
by Garcia and Chan \cite{GarciaChan99} from the experimental data
has not been understood theoretically, in particular its deep minimum
right below $\Tc$. While the value of the Casimir force at criticality
as well as the decay above $\Tc$ could be calculated using field
theory \cite{KrechDietrich92b,Krech94,DiehlGruenebergShpot06,GruenebergDiehl08},
no quantitative results were available for the scaling region $T\lesssim\Tc$
except for mean-field-theoretical approaches \cite{MaciolekGambassiDietrich07,ZandiShackellRudnickKardarChayes07}.
Analytic results exist only for the non-critical region below $\Tc$,
where contributions to the thermodynamic Casimir force from Goldstone
modes \cite{LiKardar91,LiKardar92,KardarGolestanian99} and from the
excitation of capillary waves of the liquid-vapor $^{4}$He interface
\cite{ZandiRudnickKardar04} become dominant.

This unsatisfactory situation was resolved in Ref.~\cite{Hucht07a},
where a method was proposed to calculate the thermodynamic Casimir
force for $O(n)$-symmetrical lattice models using Monte Carlo simulations
without any approximations, in contrast to, e.\,g., the stress tensor
method used by Dantchev and Krech \cite{DantchevKrech04}, which furthermore
was restricted to periodic systems. The Monte Carlo simulations were
done for the three-dimensional (3D) XY model on a simple cubic lattice
with film geometry $\Ls\times\Lp\times\Lp$ and open boundary conditions
along the $\perp$-direction, as this system is known to be in the
same universality class as the superfluid transition in $^{4}$He
and thus displays the same asymptotic critical behavior. The results
were found to be in excellent agreement with the experimental results
by Garcia, Chan and coworkers \cite{GarciaChan99,GanshinScheidemantelGarciaChan06}
and for the first time provided a theoretical explanation for the
characteristic shape of the finite-size scaling function and in particular
its deep minimum below $\Tc$. In the following, this method was used
to determine Casimir forces in various systems and geometries \cite{Hasenbusch0907,Hasenbusch1005},
while other methods for the evaluation of thermodynamic Casimir forces
using Monte Carlo simulations have also been presented \cite{VasilyevGambassiMaciolekDietrich07,Hasenbusch0905,Hasenbusch0908}.

In the present work, this method is used to derive the universal finite-size
scaling function of the excess free energy and thermodynamic Casimir
force as functions of the aspect ratio $\rho=\Ls/\Lp$ for the 3D
Ising model with cuboid geometry and periodic boundary conditions\C{,
which are in the same bulk universality class as binary liquid mixtures}.
Here $\rho$ is allowed to take arbitrary values from $\rho\to0$
(film geometry) to $\rho\to\infty$ (rod geometry), while former investigations
were either at $\rho=0$ \cite{KrechDietrich92b,Krech94,DiehlGruenebergShpot06,GruenebergDiehl08,MaciolekGambassiDietrich07,ZandiShackellRudnickKardarChayes07}
or limited to the case $\rho\ll1$ \cite{DantchevKrech04,Hucht07a,VasilyevGambassiMaciolekDietrich07,Hasenbusch0905,Hasenbusch0907,Hasenbusch0908,VasilyevGambassiMaciolekDietrich09,Hasenbusch1005,ToldinDietrich10}.
The paper is structured as follows: In the remainder of Sec.~\ref{sec:Introduction}
the basic principles and definitions concerning the thermodynamic
Casimir effect are discussed and the Monte Carlo method will be revisited.
In Sec.~\ref{sec:Results}, our Monte Carlo results are discussed
and compared to recently published results by Dohm \cite{Dohm09}
who calculated the Casimir force within a minimal renormalization
scheme of the $O(n)$ model at finite $\rho$, covering temperatures
below and above $\Tc$, as well as to field-theoretical results obtained
for $T\ge\Tc$ in the framework of the renormalization group-improved
perturbation theory (RG) to two-loop order \cite{DiehlGruenebergShpot06,GruenebergDiehl08}.
In Sec.~\ref{sec:2d-system}, we present an exact calculation of
the excess free energy and Casimir force scaling functions for the
two-dimensional (2D) Ising model with arbitrary aspect ratios $\rho$.
We conclude with a discussion and a summary.

\subsection{Basic principles}

When a thermodynamical system in $d$ dimensions such as a simple
classical fluid or a classical $n$-vector magnet is confined to a
region with thickness $\Ls$ and cross-sectional area $\Lp^{d-1}$,
its total free energy $F$ becomes explicitly size-dependent. Then
the reduced free energy per unit volume \begin{eqnarray}
f(T,\Ls,\Lp) & \equiv & \frac{F(T,\Ls,\Lp)}{\Ls\Lp^{d-1}k_{\text{B}}T}\nonumber \\
 & = & f_{\infty}(T)+\delta f(T,\Ls,\Lp)\label{eq:f_def}\end{eqnarray}
can be decomposed \cite{Privman90} into a sum of the bulk free energy
density $f_{\infty}$ and a finite-size contribution $\delta f$.
As we assume periodic boundary conditions in all directions, the surface
terms in $\perp$ and $\parallel$ directions as well as edge and
corner contributions are omitted in (\ref{eq:f_def}). In this case
the residual free energy $\delta f$ equals the excess free energy
$f_{\mathrm{ex}}$, and we will use $f_{\mathrm{ex}}$ instead of
$\delta f$ in the following.

In terms of $f_{\mathrm{ex}}$ the reduced thermodynamic Casimir force
per surface area in $\perp$ direction is defined as \cite{Krech94}
\begin{equation}
\Fc(T,\Ls,\Lp)\equiv-\frac{\partial[\Ls f_{\mathrm{ex}}(T,\Ls,\Lp)]}{\partial\Ls},\label{eq:FC_def}\end{equation}
where $\beta=1/k_{\text{B}}T$, and the derivative is taken at fixed
$\Lp$. We omit the index $\perp$ for the Casimir force, as we will
not consider Casimir forces in parallel directions in this work.

When in the absence of symmetry breaking external fields the critical
point is approached from higher temperatures, which corresponds to
the liquid-vapor critical point in the case of a simple classical
fluid or to the Curie point in a classical $n$-vector magnet, the
bulk correlation length $\xi_{\infty}(t)$ grows and diverges as %
\footnote{Throughout this work, the symbol $\sim$ means {}``asymptotically
equal'' in the respective limit, $\Lp,\Ls\rightarrow\infty$, $T\rightarrow\Tc$,
keeping the scaling variables $x$ and $\rho$ fixed, i.\,e., $f(L)\sim g(L)\Leftrightarrow\lim_{L\rightarrow\infty}f(L)/g(L)=1.$%
} \begin{equation}
\xi_{\infty}(t)\underset{t>0}{\sim}\xip t^{-\nu},\label{eq:xi_def}\end{equation}
with correlation length exponent $\nu$, reduced temperature $t=T/T_{\text{c},\infty}-1$,
and non-universal amplitude $\xip$. In this work we use $\xip=Q_{\xi}^{+}f^{+}=1.000183(2)\times0.506(1)$
valid for the 3D Ising model on a simple cubic lattice \cite{CampostriniPelissettoRossiVicari99,ButeraComi02}. 

According to the theory of finite-size scaling \cite{Fisher71} and
under the assumption, that long-range interactions and other contributions
irrelevant in the RG sense are negligible, as for instance subleading
long-range interactions \cite{DantchevDiehlGrueneberg06}, the thermodynamic
Casimir force in the regime $\Ls,\Lp,\xi_{\infty}\gg a$, where $a$
is a characteristic microscopic length scale such as the lattice constant
in the case of a lattice model, obeys a finite-size scaling form \begin{equation}
\Fc(T,\Ls,\Lp)\sim\Ls^{-d}\Xc_{\perp}(x_{\perp},\rho),\label{eq:FSS_FC_s}\end{equation}
where the scaling variable $x_{\perp}$ can be defined as \begin{equation}
x_{\perp}\equiv t\left(\frac{\Ls}{\xip}\right)^{\frac{1}{\nu}}\underset{t>0}{\sim}\left(\frac{\Ls}{\xi_{\infty}(t)}\right)^{\frac{1}{\nu}},\label{eq:x_s}\end{equation}
$\rho=\Ls/\Lp$ denotes the aspect ratio, and $\Xc_{\perp}$ is a\emph{
}finite-size scaling function. Note that in this work $\Xc$ always
denotes the scaling function of the Casimir force in $\perp$ direction,
while the index describes the reference direction $\perp$ or $\parallel$
of length $L$.

An analogous finite-size scaling relation holds for the excess free
energy, \begin{equation}
f_{\mathrm{ex}}(T,\Ls,\Lp)\sim\Ls^{-d}\Theta_{\perp}(x_{\perp},\rho),\label{eq:FSS_fex_s}\end{equation}
and $\Xc_{\perp}$ is related to $\Theta_{\perp}$ according to \cite{Dohm09}\begin{equation}
\Xc_{\perp}(x_{\perp},\rho)=\left[d-1-\frac{1}{\nu}\frac{x_{\perp}\partial}{\partial x_{\perp}}-\frac{\rho\partial}{\partial\rho}\right]\Theta_{\perp}(x_{\perp},\rho).\label{eq:ScalingIdentity_s}\end{equation}
The dimensionless finite-size scaling functions $\Theta_{\perp}$
and $\Xc_{\perp}$ are universal, that is, they only depend on gross
properties of the system such as the bulk and surface universality
classes of the phase transition, the system shape and boundary conditions,
but not on its microscopic details \cite{GruenebergHucht04,DantchevDiehlGrueneberg06}.

At the critical point $\Tc$ the thermodynamic Casimir force becomes
long-ranged and for sufficiently large values of the length $\Ls$
asymptotically decays as \begin{eqnarray}
\Fc(\Tc,\Ls,\Lp) & \sim & L_{\perp}^{-d}\Xc_{\perp}(0,\rho)\nonumber \\
 & \sim & L_{\perp}^{-d}[(d-1)\Dcs(\rho)-\rho\Dcs'(\rho)],\label{eq:FSS_FC_s_Tc}\end{eqnarray}
where $\Dcs(\rho)\equiv\Theta_{\perp}(0,\rho)$ is the so-called Casimir
amplitude \cite{FisherdeGennes78}, being -- like the finite-size
scaling function $\Xc_{\perp}$ -- an universal quantity. Note that
for finite aspect ratios $\rho>0$ the Casimir amplitude becomes $\rho$-dependent.
The film geometry is recovered by letting $\rho\rightarrow0$, and
Eq.~(\ref{eq:FSS_FC_s_Tc}) simplifies to \begin{equation}
\Fc(\Tc,\Ls,\infty)\sim L_{\perp}^{-d}(d-1)\Dcs(0).\label{eq:FSS_FC_s_Tc_0}\end{equation}

Since the 1990s, such universal quantities have been subject of extensive
theoretical research. They were studied by means of exactly solvable
models \cite{Krech94,Dantchev98,BrankovDantchevTonchev00,DantchevKrechDietrich03,DantchevKrech04,DantchevDiehlGrueneberg06,DantchevGrueneberg09},
Monte Carlo simulations \cite{KrechLandau96,DantchevKrech04,Hucht07a,VasilyevGambassiMaciolekDietrich07,MaciolekGambassiDietrich07,HertleinHeldenGambassiDietrichBechinger08,Hasenbusch0905,Hasenbusch0907,Hasenbusch0908,VasilyevGambassiMaciolekDietrich09,Hasenbusch1005},
as well as within field-theoretical approaches \cite{KrechDietrich92b,Krech94,DiehlGruenebergShpot06,GruenebergDiehl08,SchmidtDiehl08,DiehlGrueneberg09,Dohm09,BurgsmuellerDiehlShpot10}.

\subsection{Reformulation for arbitrary $\rho$}

The formulation of the Casimir force finite-size scaling laws in the
previous section was done by assuming film geometry $\rho\ll1$, i.\,e.,
having in mind the limit $\rho\rightarrow0$. However, if $\rho\gtrsim1$
this picture is not appropriate and should be replaced by a more general
treatment. In the following we rewrite the basic scaling laws in terms
of the system volume $V=\Ls\Lp^{d-1}$ instead of the film thickness
$\Ls$. The resulting scaling functions can be used in the whole regime
$0<\rho<\infty$.

Using the substitution $\Ls^{d}\to V\rho^{d-1}$ in Eq.~(\ref{eq:FSS_fex_s})
we get\begin{equation}
f_{\mathrm{ex}}(T,\Ls,\Lp)\sim V^{-1}\Theta(x,\rho)\label{eq:FSS_fex}\end{equation}
with an universal scaling function $\Theta$ and the generalized scaling
variable \begin{equation}
x\equiv t\left(\frac{V}{\xip^{d}}\right)^{\frac{1}{d\nu}},\label{eq:x}\end{equation}
while the scaling function $\Theta_{\perp}$ from Eq.~(\ref{eq:FSS_fex_s})
is recovered as \begin{equation}
\Theta_{\perp}(x_{\perp},\rho)=\rho^{d-1}\Theta(x,\rho),\label{eq:Theta_s}\end{equation}
with \begin{equation}
x_{\perp}=\rho^{\frac{1}{\nu}-\frac{1}{d\nu}}x.\label{eq:x_s(x)}\end{equation}
Similarly, the Casimir force obeys \begin{equation}
\Fc(T,\Ls,\Lp)\sim V^{-1}\Xc(x,\rho),\label{eq:FSS_FC}\end{equation}
 from which we derive the scaling identity\begin{equation}
\Xc(x,\rho)=-\left[\frac{1}{d\nu}\frac{x\partial}{\partial x}+\frac{\rho\partial}{\partial\rho}\right]\Theta(x,\rho).\label{eq:ScalingIdentity}\end{equation}
Note that this identity is equivalent to but simpler than Eq.~(\ref{eq:ScalingIdentity_s}).
At criticality we now define the generalized Casimir amplitude as
\begin{equation}
\Dc(\rho)=\Theta(0,\rho)\label{eq:Delta}\end{equation}
 and find \begin{eqnarray}
f_{\mathrm{ex}}(\Tc,\Ls,\Lp) & \sim & V^{-1}\Dc(\rho),\label{eq:fex(Delta)}\\
\Fc(\Tc,\Ls,\Lp) & \sim & -V^{-1}\rho\Dc'(\rho).\label{eq:FC(Delta)}\end{eqnarray}

The case $\rho=1$ deserves special attention: As\begin{equation}
\left.\frac{\partial}{\partial\rho}\Theta(x,\rho)\right|_{\rho=1}=0\label{eq:dTheta-drho.eq.0}\end{equation}
in cubic geometry (see Appendix~\ref{sec:Stationarity-of-Theta}),
Eq.~(\ref{eq:ScalingIdentity}) simplifies to \begin{equation}
\Xc(x,1)=-\frac{1}{d\nu}\frac{x\partial}{\partial x}\Theta(x,1)\label{eq:ScalingIdentity_rho.eq.1}\end{equation}
at $\rho=1$, and gives a remarkably simple connection between the
Casimir force and the excess internal energy, Eq.~(\ref{eq:uex}),
in the cube shaped system, namely\begin{equation}
\Fc(T,L,L)\sim\frac{t}{d\nu}\, u_{\mathrm{ex}}(T,L,L).\label{eq:FC_von_uex}\end{equation}
Obviously, the Casimir force vanishes at the critical point if $\rho=1$,
i.\,e., \begin{equation}
\Xc(0,1)=0.\label{eq:theta(0,1)}\end{equation}

For completeness we also give the definitions of the scaling functions
in terms of $\Lp$. As \begin{equation}
f_{\mathrm{ex}}(T,\Ls,\Lp)\sim\Lp^{-d}\Theta_{\parallel}(x_{\parallel},\rho),\label{eq:FSS_fex_p}\end{equation}
we find \begin{equation}
\Theta_{\parallel}(x_{\parallel},\rho)=\rho^{-1}\Theta(x,\rho)\label{eq:Theta_p}\end{equation}
with $x_{\parallel}\equiv t(\Lp/\xip)^{1/\nu}$. Note that in this
representation the scaling identity Eq.~(\ref{eq:ScalingIdentity_s})
reads\begin{equation}
\Xc_{\parallel}(x_{\parallel},\rho)=-\left[1+\frac{\rho\partial}{\partial\rho}\right]\Theta_{\parallel}(x_{\parallel},\rho)\label{eq:ScalingIdentity_p}\end{equation}
 and in the limit $\rho\to\infty$ simplifies to \begin{equation}
\Xc_{\parallel}(x_{\parallel},\infty)=-\Theta_{\parallel}(x_{\parallel},\infty),\label{eq:ScalingIdentity_inf}\end{equation}
leading to the simple relation \begin{equation}
\Fc(T,\infty,\Lp)\sim-f_{\mathrm{ex}}(T,\infty,\Lp).\label{eq:FC(T,inf,Lp)}\end{equation}

\subsection{Monte Carlo method}

\begin{figure}
\begin{centering}
\includegraphics[scale=0.6]{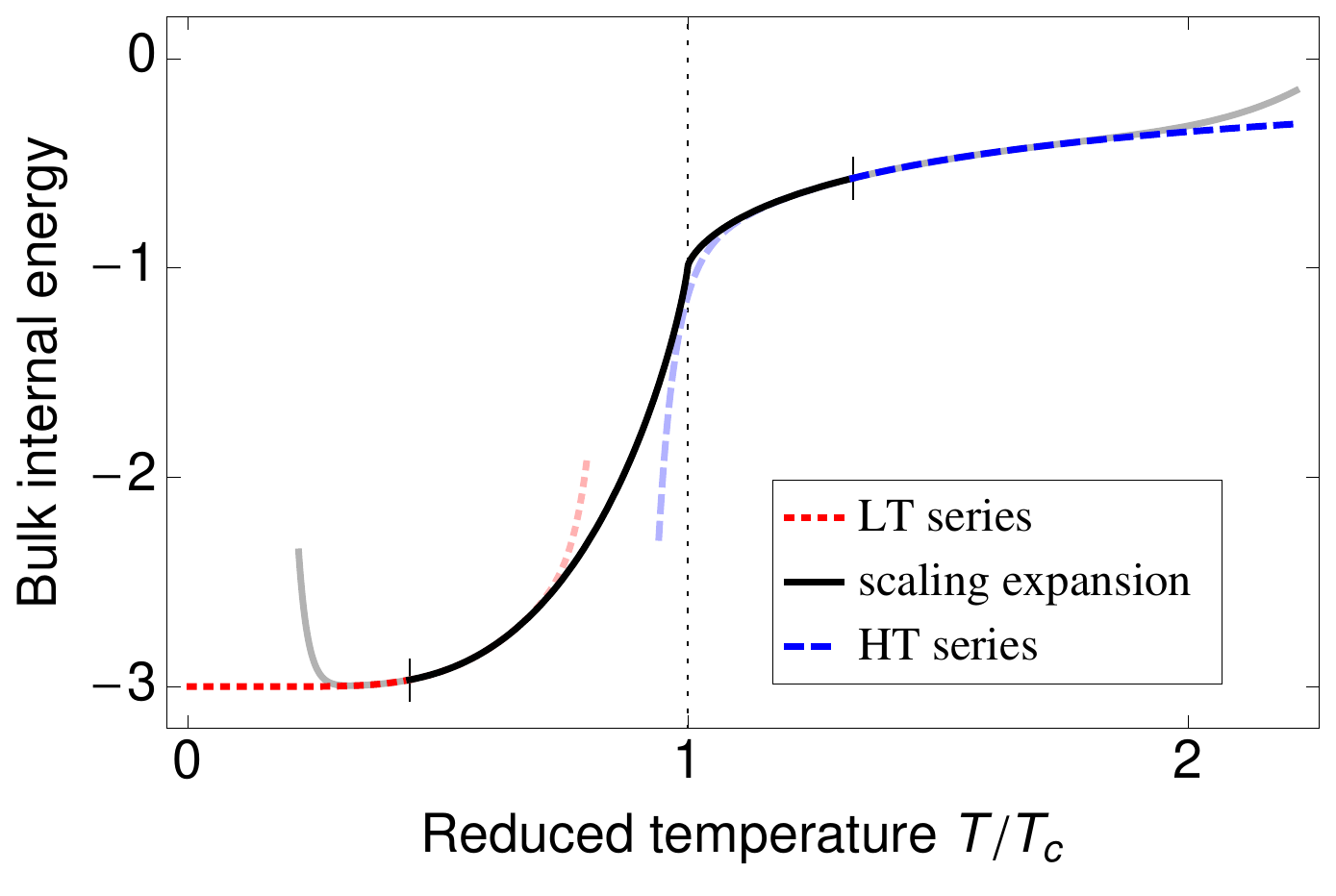}
\par\end{centering}

\caption{(Color online) Bulk internal energy density $e_{\infty}(T)$ obtained
from three different methods: low temperature series \cite{BhanotCreutzHorvathLackiWeckel94}
(red dotted line), scaling expansion \cite{FengBloete10} (black line),
and high temperature series \cite{ArisueFujiwara03a} (blue dashed
line). \label{fig:Einf}}

\end{figure}
In this work we focus on the three-dimensional isotropic nearest neighbor
Ising model on a $\Ls\times\Lp\times\Lp$ simple cubic lattice with
periodic boundary conditions and Hamiltonian \begin{equation}
\beta\mathcal{H}=-\frac{K}{2}\sum_{\langle ij\rangle}\sigma_{i}\sigma_{j},\label{eq:Hamiltonian}\end{equation}
where $K=\beta J>0$ is the ferromagnetic reduced exchange interaction
and $\sigma_{i}=\pm1$ are one-component spin variables at lattice
sites $i$. The Monte Carlo simulations were done using the Wolff
single cluster algorithm \cite{Wolff89}. Measuring the reduced internal
energy density \begin{equation}
u(T,\Ls,\Lp)=\frac{\langle\beta\mathcal{H}\rangle}{\Lp^{d-1}\Ls},\label{eq:u}\end{equation}
the excess free energy and Casimir force is calculated as follows
\cite{Hucht07a}: First we determined the excess internal energy \begin{equation}
u_{\text{ex}}(T,\Ls,\Lp)\equiv u(T,\Ls,\Lp)-u_{\infty}(T)\label{eq:uex}\end{equation}
by subtracting the reduced bulk internal energy density $u_{\infty}(T)$.
We used three different results to get precise estimates for $u_{\infty}(T)$
of the $3d$ Ising model in the different temperature regimes: For
low temperatures $K>1/2$ we used the low temperature series expansion
(54th order) by Bhanot \emph{et}\,\emph{al.} \cite{BhanotCreutzHorvathLackiWeckel94},
while for $K<1/6$ the high temperature series expansion (46th order)
by Arisue and Fujiwara \cite{ArisueFujiwara03a} was utilized. Finally,
in the vicinity of the critical point we used the expansion recently
obtained by Feng and Blöte \cite{FengBloete10}, where we also took
the bulk critical indices \cite{DengBloete03} $K_{\mathrm{c}}=0.22165455(3)$,
$\nu^{-1}=1.5868(3)$ and $\omega=0.821(5)$. These three estimates
of $u_{\infty}(T)$ show a broad overlap, see also the discussion
by Feng and Blöte \cite{FengBloete10}, the resulting non-reduced
bulk internal energy density $e_{\infty}(T)=u_{\infty}(T)/\beta$
is depicted in Fig.~\ref{fig:Einf}.\textcolor{red}{{} }With the identity\begin{equation}
\fex(T,\Ls,\Lp)=-\int_{T}^{\infty}\frac{\mathrm{d}\tau}{\tau}u_{\mathrm{ex}}(\tau,\Ls,\Lp)\label{eq:fex_integral}\end{equation}
we determined $\fex$ by numerical integration, using the fact that
$u_{\mathrm{ex}}$ goes exponentially fast to zero above $\Tc$ \cite{Hucht07a}. 

To obtain the Casimir force, we first calculated the \emph{internal
Casimir force} 

\begin{equation}
\Fi(T,\Ls,\Lp)=-\frac{\partial[\Ls u_{\text{ex}}(T,\Ls,\Lp)]}{\partial\Ls},\label{eq:FI}\end{equation}
which is defined similar to Eq.~(\ref{eq:FC_def}), by numerical
differentiation, using thicknesses $\Ls'=\Ls\pm1$ in order to get
an integral effective thickness $\Ls$. With Eqs.~(\ref{eq:FSS_FC_s},
\ref{eq:FC_integral}) and the hyperscaling relation $d\nu=2-\alpha$
with specific heat exponent $\alpha$, it is straightforward to show
that this quantity fulfills the finite-size scaling form\begin{equation}
-\Fi(T,\Ls,\Lp)\sim\xip^{-1/\nu}\Ls^{(\alpha-1)/\nu}\Xc_{\perp}'(x_{\perp},\rho),\label{eq:X_internal}\end{equation}
with an universal finite-size scaling function \begin{equation}
\Xc_{\perp}'(x_{\perp},\rho)=\frac{\partial\Xc_{\perp}(x_{\perp},\rho)}{\partial x_{\perp}}.\label{eq:theta_s'}\end{equation}
This quantity turns out to be very useful in understanding the Casimir
force scaling function $\Xc_{\perp}(x_{\perp},\rho)$ for $\rho\to0$,
as will be shown in the next section. Finally, the thermodynamic Casimir
force is obtained from Eq.~(\ref{eq:FI}) by integration, \begin{equation}
\Fc(T,\Ls,\Lp)=-\int_{T}^{\infty}\frac{\mathrm{d}\tau}{\tau}\Fi(\tau,\Ls,\Lp),\label{eq:FC_integral}\end{equation}
 where again the exponential decay above $\Tc$ simplifies the numerical
integration.

\section{Results\label{sec:Results}}

\subsection{Casimir force in film geometry $\rho\to0$ \label{sub:rho.to.0}}

\begin{figure}
\begin{centering}
\includegraphics[scale=0.6]{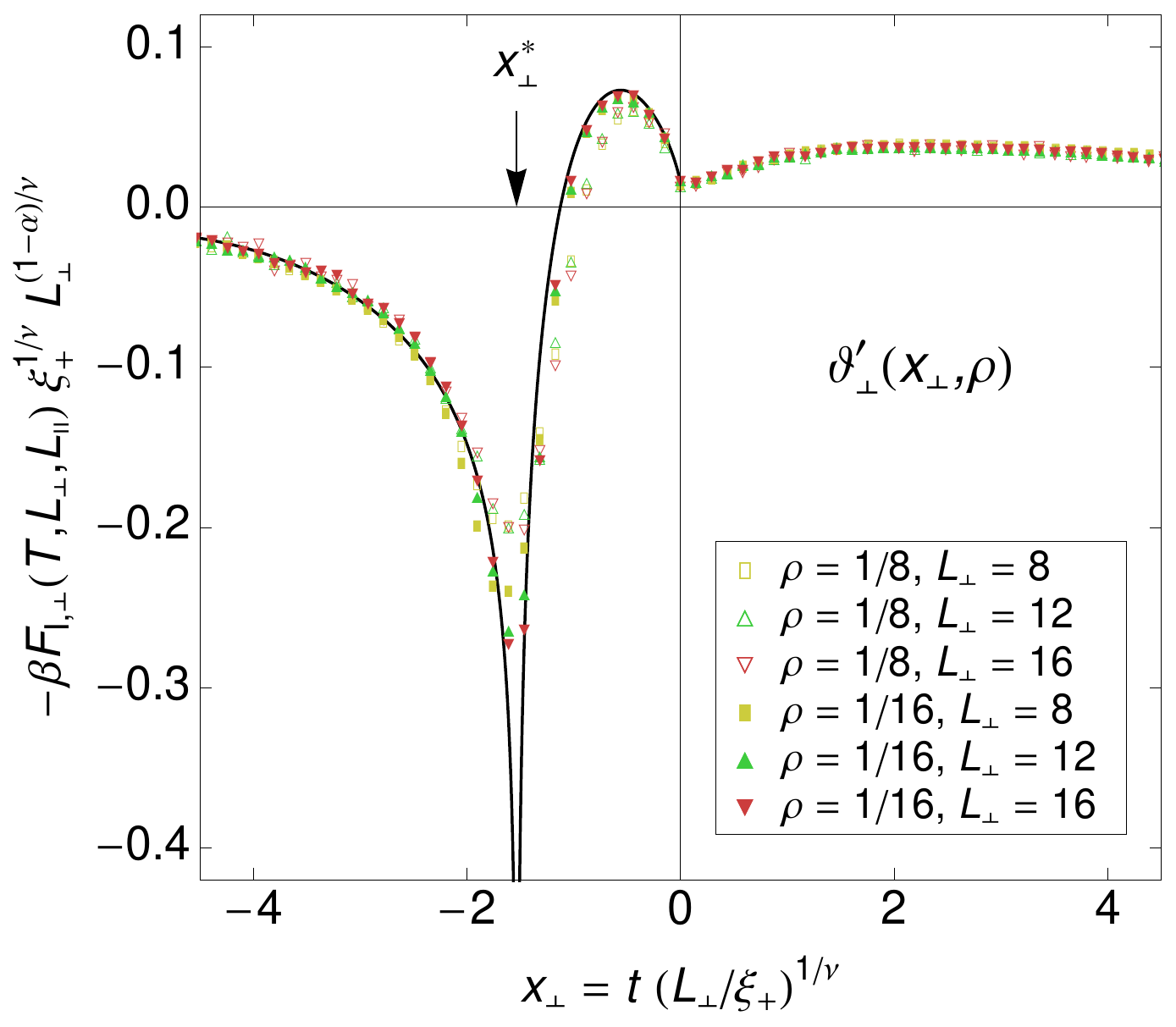}
\par\end{centering}

\caption{(Color online) Internal Casimir force scaling function $\Xc_{\perp}'(x_{\perp},\rho)$
for small aspect ratios $\rho=1/16$ and $\rho=1/8$. The black thin
line is the extrapolation $\rho\to0$, showing a logarithmic singularity
at $x_{\perp}^{*}=1.535(10)$ (see text). \label{fig:Fint1}}

\end{figure}

In Fig.~\ref{fig:Fint1} we plot the internal Casimir force $\Fi$,
Eq.~(\ref{eq:FI}), for small aspect ratios $\rho=1/8$ and $\rho=1/16$.
In the limit of film geometry $\rho\to0$ we observe strong finite-size
effects below the critical point \cite{Hucht07a}, which are caused
by the influence of the phase transition in the $d{-}1$-dimensional
system. In this section we will analyse this influence in detail and
show that $\Fi$ is directly connected to the specific heat of the
$d{-}1$-dimensional system. We will give the derivation for periodic
systems where no surface terms occur, as these terms will complicate
the analysis \cite{Hasenbusch0907}.

From the scaling identity Eq.~(\ref{eq:ScalingIdentity_s}) for $\rho\to0$,\begin{equation}
\Xc_{\perp}(x_{\perp},0)=\left[d-1-\frac{1}{\nu}\frac{x_{\perp}\partial}{\partial x_{\perp}}\right]\Theta_{\perp}(x_{\perp},0)\label{eq:theta_s(x_s,0)}\end{equation}
we get \begin{equation}
\Fc\underset{\rho\to0}{\sim}(d-1)f_{\mathrm{ex}}+t\,\nu^{-1}\, u_{\mathrm{ex}},\label{eq:FC_rho.eq.0}\end{equation}
i.\,e., within the scaling region and for $\rho\text{\ensuremath{\to}0 }$
the Casimir force can alternatively be calculated without $\Ls$-derivative
\cite{Hasenbusch0907}. For the internal Casimir force scaling function\begin{equation}
\Xc_{\perp}'(x_{\perp},0)=\left[d-1-\frac{1}{\nu}-\frac{1}{\nu}\frac{x_{\perp}\partial}{\partial x_{\perp}}\right]\frac{\partial\Theta_{\perp}(x_{\perp},0)}{\partial x_{\perp}}\label{eq:theta_s^prime}\end{equation}
 we find the asymptotic identity \begin{equation}
-\Fi\underset{\rho\to0}{\sim}-\left[d-1-\frac{1-t}{\nu}\right]\, u_{\mathrm{ex}}+\frac{t}{\nu}\, c_{\mathrm{ex}}\label{eq:FI_rho.eq.0}\end{equation}
with the excess specific heat\begin{equation}
c_{\mathrm{ex}}(T,\Ls,\Lp)\equiv c(T,\Ls,\Lp)-c_{\infty}(T)\label{eq:c_ex}\end{equation}
and $c=\partial Tu/\partial T$ as usual. For $\rho\to0$, this quantity
contains both the bulk singularity \begin{equation}
c_{\infty}(T)\sim A_{\pm}\left|t\right|^{-\alpha},\label{eq:c_bulk}\end{equation}
with amplitudes $A_{\pm}$, as well as the singularity of the laterally
infinite film with finite thickness $\Ls$ at $t_{\mathrm{c}}(\Ls)=T_{\mathrm{c}}(\Ls)/\Tc-1$,
which scales as \begin{equation}
c(T,\Ls,\infty)\sim A_{\pm}^{*}\left(\frac{\Ls}{\xip}\right)^{\frac{\alpha-\alpha^{*}}{\nu}}\left|t-t_{\mathrm{c}}(\Ls)\right|^{-\alpha^{*}}.\label{eq:c_film}\end{equation}
Here, $\alpha^{*}$ denotes the specific heat exponent of the $d{-}1$-dimensional
system, $A_{\pm}^{*}$ are amplitudes, and the factor $(\Ls/\xip)^{(\alpha-\alpha^{*})/\nu}$
guarantees the correct scaling behavior for $\Ls\to\infty$ by cancellation
of terms containing $\alpha^{*}$. 

However, as $c_{\mathrm{ex}}$ enters Eq.~(\ref{eq:FI_rho.eq.0})
with prefactor $t$ only, the bulk singularity at $t=0$ is suppressed
(as $\alpha<1$) and $c_{\mathrm{ex}}$ is dominated by the singularity
from Eq.~(\ref{eq:c_film}), at\begin{equation}
x_{\perp}^{*}\sim t_{\mathrm{c}}(\Ls)\,\left(\frac{\Ls}{\xip}\right)^{\frac{1}{\nu}}.\label{eq:x_s^star}\end{equation}
The location of the critical point was re-analysed from the data of
Kitatani \emph{et}\,\emph{al.} \cite{KitataniOhtaIto96} including
corrections to scaling, as well as from the data of Caselle and Hasenbusch
\cite{CaselleHasenbusch96}, giving the value $x_{\perp}^{*}=-1.535(10)$.
This improves the value $x_{\perp}^{*}=-1.60(2)$ found by Vasilyev
\emph{et}\,\emph{al.} \cite{VasilyevGambassiMaciolekDietrich09}.
Furthermore, the other terms in (\ref{eq:FI_rho.eq.0}) are $O(1)$
near $x_{\perp}^{*}$, which leads us to the conclusion that the specific
heat singularity of the $d{-}1$-dimensional film is directly visible
in the scaling function $\Xc_{\perp}'(x_{\perp},0)$ around $x_{\perp}=x_{\perp}^{*}$,\begin{equation}
\Xc_{\perp}'(x_{\perp}\approx x_{\perp}^{*},0)\sim\frac{A_{\pm}^{*}\xip^{d}}{\nu}\, x_{\perp}|x_{\perp}-x_{\perp}^{*}|^{-\alpha^{*}}+\mathcal{O}(1).\label{eq:theta(x*)}\end{equation}
From this arguments we conclude that the scaling function $\Xc_{\perp}'(x_{\perp},0)$
has a singularity at $x_{\perp}^{*}$ dominated by the specific heat
singularity of the $d{-}1$-dimensional system, with critical exponent
$\alpha^{*}$. In our case, $\alpha^{*}=0$ and the singularity is
logarithmic. This asymptotic behavior is included in Fig.~\ref{fig:Fint1}
as solid line. 

\begin{figure}
\begin{centering}
\includegraphics[scale=0.6]{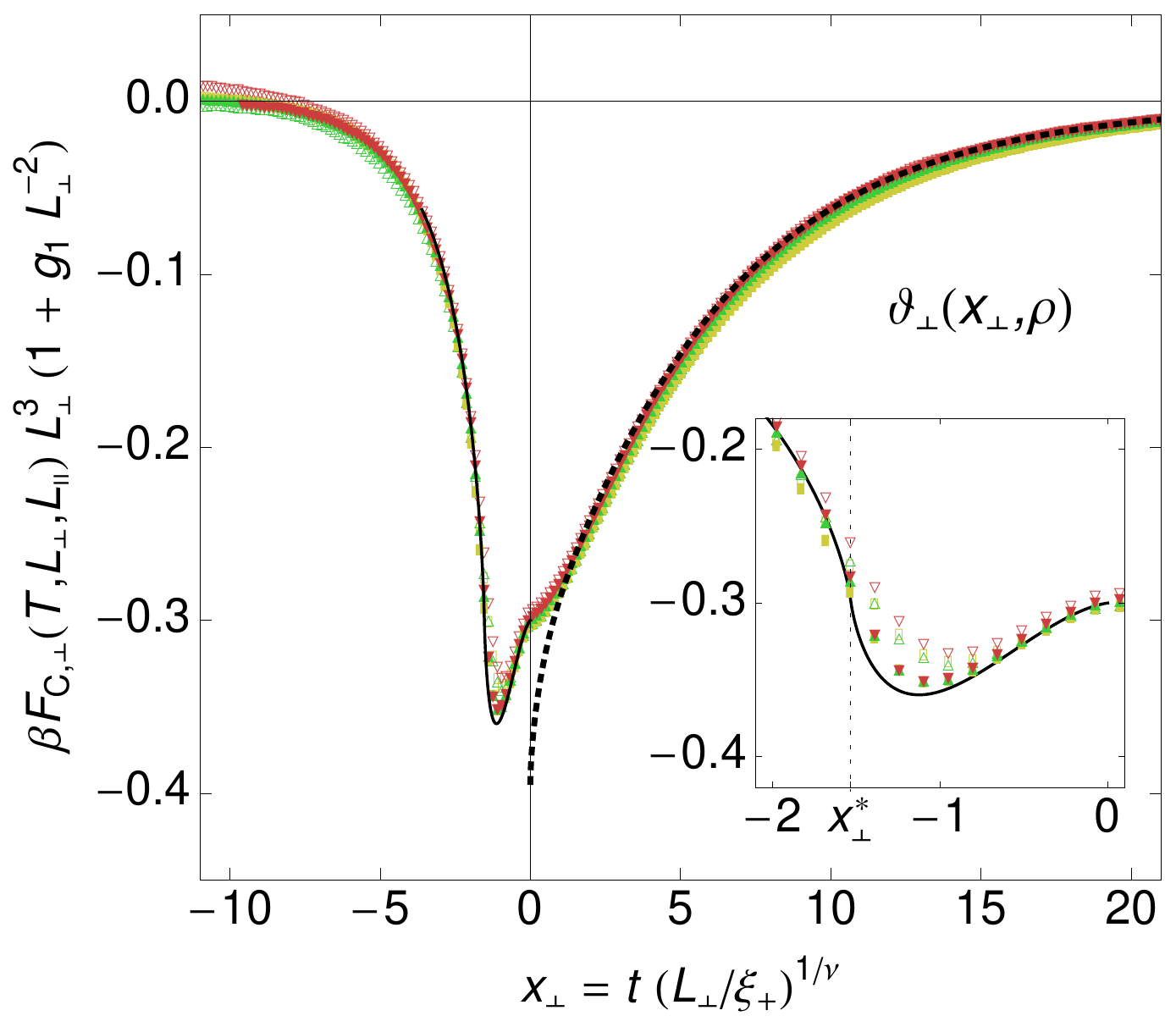}
\par\end{centering}

\caption{(Color online) Casimir force scaling function $\Xc_{\perp}(x_{\perp},\rho)$
for small aspect ratios $\rho=1/16$ and $\rho=1/8$. The solid line
is the extrapolation $\rho\to0$ calculated from the integrated logarithmic
singularity in $\Xc_{\perp}'(x_{\perp},0)$. The dotted line is the
RG calculation of Grüneberg and Diehl \cite{GruenebergDiehl08}. \label{fig:FC1}}

\end{figure}

In Fig.~\ref{fig:FC1} we show the scaling function of the Casimir
force for $\rho=1/8,1/16$, together with the RG results of Grüneberg
and Diehl \cite{GruenebergDiehl08}. The solid line is the integrated
extrapolation discussed above. We used a correction factor $(1+g_{1}\Ls^{-2})$,
with $g_{1}=-4(1)$, to account for leading systematic errors from
the discrete derivative, which are expected to be $\propto\Ls^{-2}$
in periodic systems. The inset is a magnification of the minimum,
from the divergence of $\Xc_{\perp}'(x_{\perp}{=}x_{\perp}^{*},0)$
$ $ the slope of $\Xc_{\perp}(x_{\perp},0)$ at $x_{\perp}^{*}$
diverges logarithmically. We find a critical amplitude $\Xc_{\perp}(0,0)=-0.310(6)$
(see Tab.~\ref{tab:Casimir-amplitudes}), which agrees within error
bars with the values $\Xc_{\perp}(0,0)=-0.3040(4)$ \cite{VasilyevGambassiMaciolekDietrich09}
as well as $\Xc_{\perp}(0,0)=-0.3052(20)$ \cite{Krech97}. The zero
at $\Xc_{\perp}'(x_{\perp}^{\mathrm{min}},0)$ (solid line in Fig.~\ref{fig:Fint1})
gives the minimum position $x_{\perp}^{\mathrm{min}}=-1.13(5)$, with
$\Xc_{\perp}(x_{\perp}^{\mathrm{min}},0)=-0.360(5)$, while the finite
$\rho$ results are $\Xc_{\perp}(x_{\perp}^{\mathrm{min}}=-1.10(5),1/16)=-0.352(5)$
and $\Xc_{\perp}(x_{\perp}^{\mathrm{min}}=-0.95(5),1/8)=-0.340(5)$.

\subsection{Casimir force for finite $\rho$}

\begin{figure*}
\begin{centering}
\includegraphics[scale=0.6]{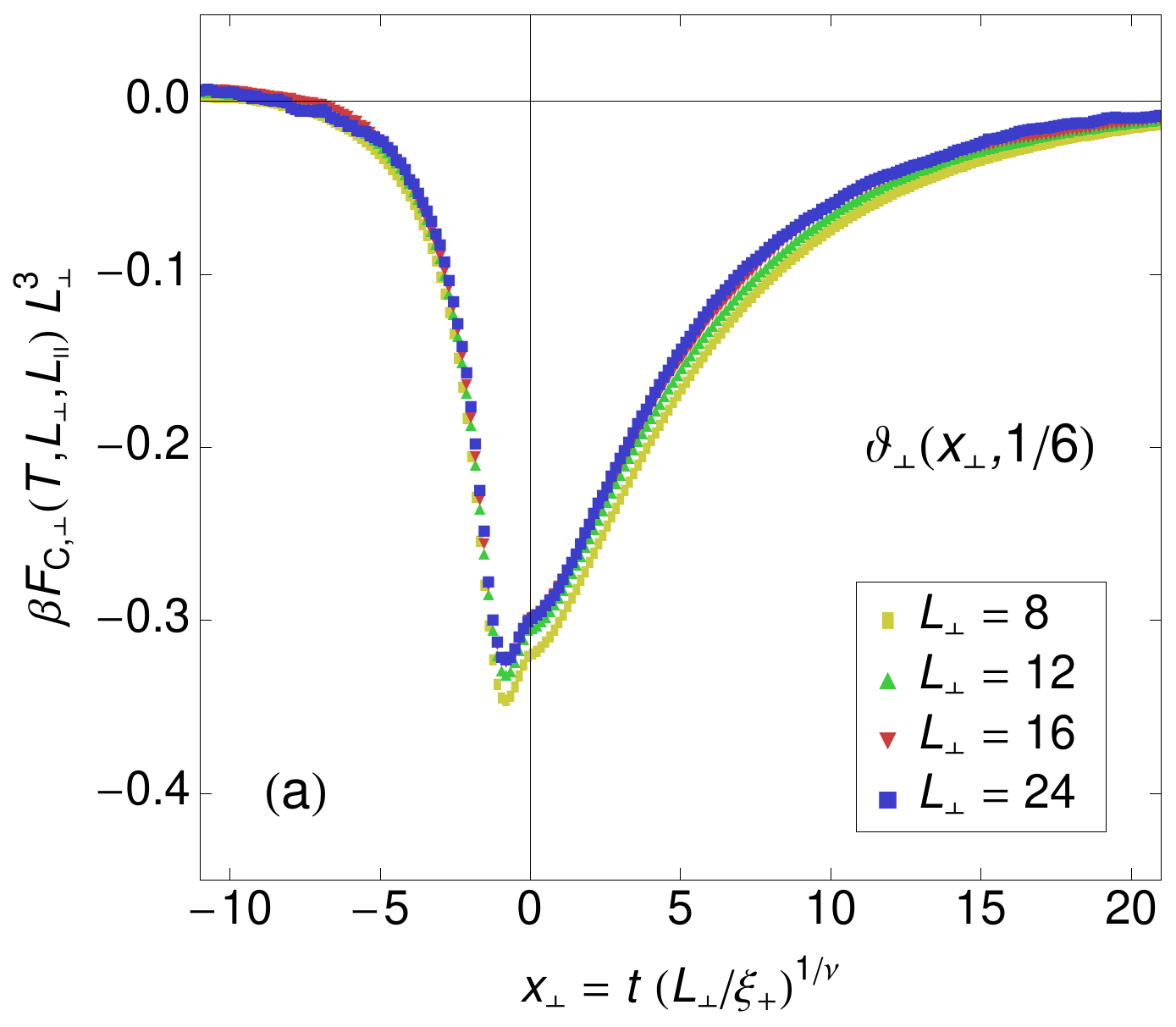}\hfill{}\includegraphics[scale=0.6]{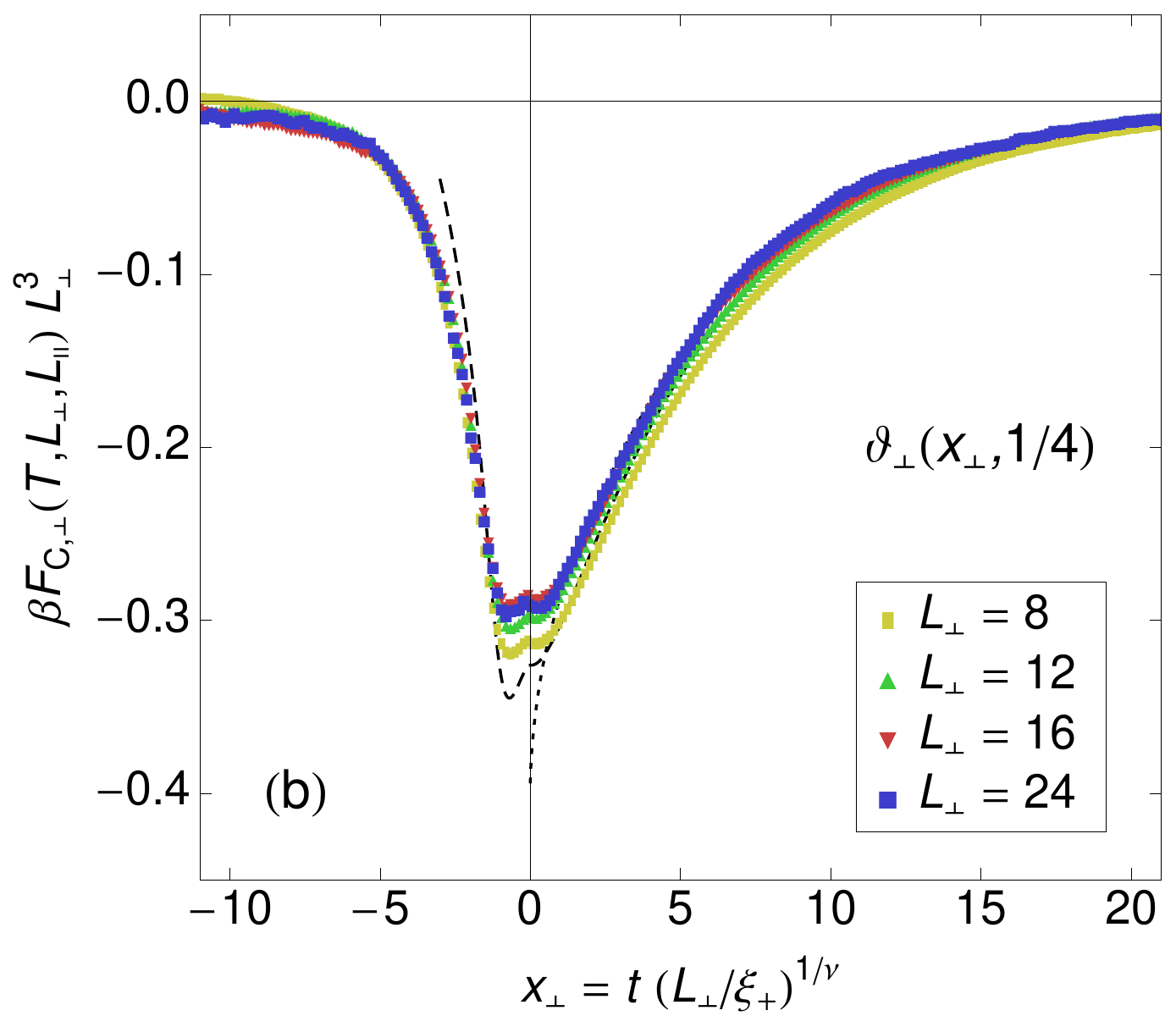}
\par\end{centering}

\begin{centering}
\includegraphics[scale=0.6]{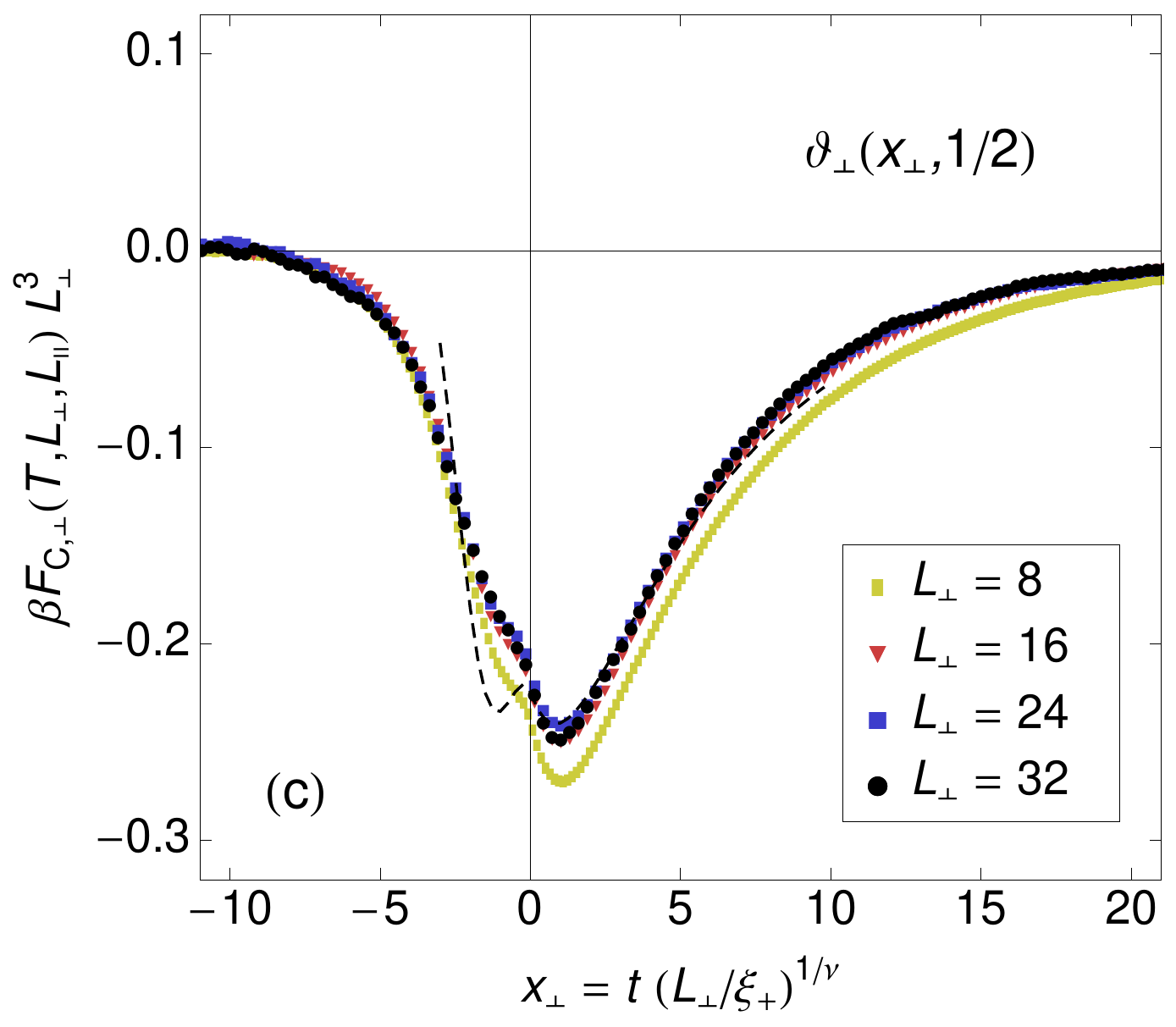}\hfill{}\includegraphics[scale=0.6]{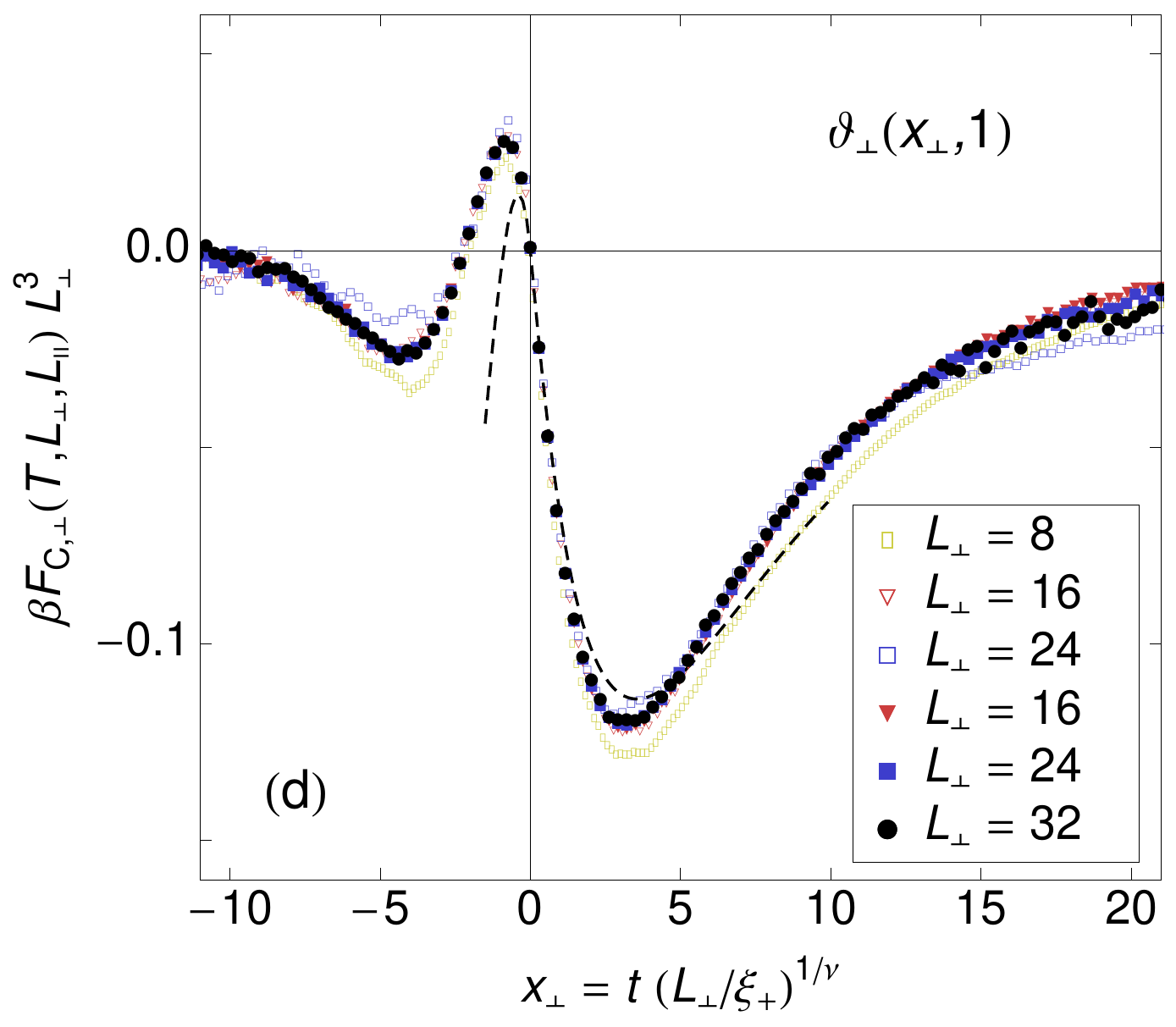}
\par\end{centering}

\caption{(Color online) Casimir force scaling function $\Xc_{\perp}(x_{\perp},\rho)$
for several aspect ratios $\rho=\{1/6,1/4,1/2,1\}$. The dotted line
is the result of Grüneberg and Diehl \cite{GruenebergDiehl08} for
$\rho=0$, while the dashed lines are the predictions of Dohm \cite{Dohm09}.
For $\rho=1$ we also show results from scaling relation Eq.~(\ref{eq:ScalingIdentity_rho.eq.1})
(filled symbols), which have much better statistics, as they are directly
calculated from the internal energy. \label{fig:CasimirPlots}}

\end{figure*}
\begin{figure}
\begin{centering}
\includegraphics[scale=0.6]{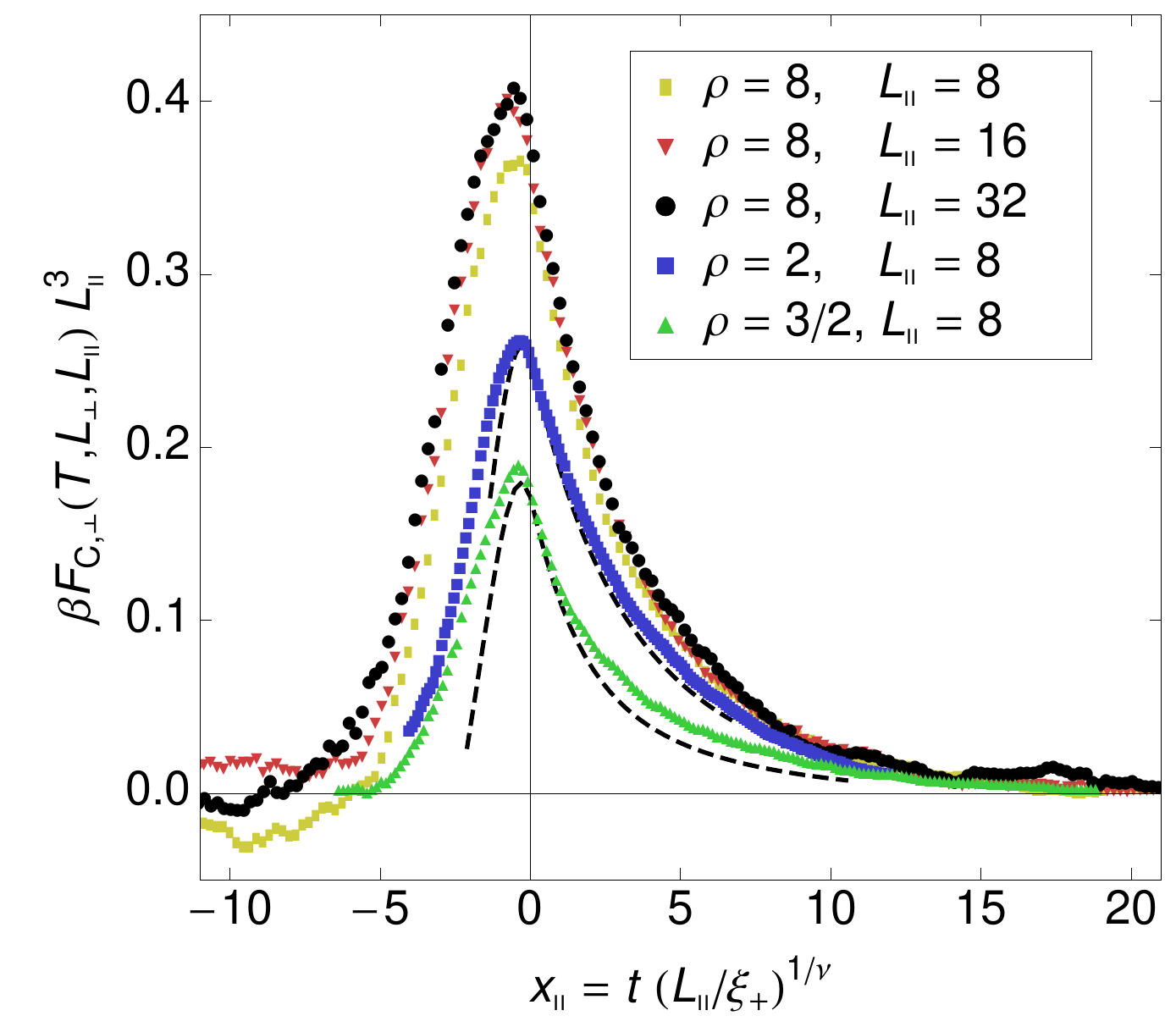}
\par\end{centering}

\caption{(Color online) Casimir force scaling function $\Xc_{\parallel}(x_{\parallel},\rho)$
for aspect ratios $\rho>1$, now as function of the proper scaling
variable $x_{\parallel}$. The dashed lines are the predictions of
Dohm \cite{Dohm09}. \label{fig:theta_p}}

\end{figure}
If we increase $\rho$ to finite values, the Casimir force scaling
function $\Xc_{\perp}(x_{\perp},\rho)$ first changes its shape around
the minimum. The results for $\rho=1/6$ (Fig.~\ref{fig:CasimirPlots}a)
already deviate distinctly from the thinner systems, the minimum below
$\Tc$ is not so deep anymore, with $\Xc_{\perp}(x_{\perp}^{\mathrm{min}}=-0.77(5),1/6)=-0.323(5)$.
These values deviate only slightly from the results of Vasilyev \emph{et}\,\emph{al.}
\cite{VasilyevGambassiMaciolekDietrich09}, $x_{\perp}^{\mathrm{min}}=-0.681(1)$
and $\Xc_{\perp}(x_{\perp}^{\mathrm{min}},1/6)=-0.329(1)$, which
we attribute to the larger statistical error in Ref.~\cite{VasilyevGambassiMaciolekDietrich09}.

When the aspect ratio is further increased to $\rho=1/4$ (Fig.~\ref{fig:CasimirPlots}b),
the curve has two minima below and above $\Tc$ which are nearly equal
in depth. Note that the results for $\rho\geq1/4$ are compared to
the predictions of Dohm \cite{Dohm09} and show similar behavior.
For $\rho\gtrsim1/4$ the minimum below $\Tc$ vanishes, while the
one above $\Tc$ remains. This is shown in Fig.~\ref{fig:CasimirPlots}c,
where we plot the Casimir scaling function for $\rho=1/2$.

The results for the cube shaped system with $\rho=1$ are shown in
Fig.~\ref{fig:CasimirPlots}d %
\footnote{At $\rho=1$ all quantities $q$ obey $q=q_{\perp}=q_{\parallel}$ %
}. The case $\rho=1$ is quite interesting, as here the Casimir force
at $x=0$ vanishes (Eq.~(\ref{eq:theta(0,1)})) and even becomes
positive for $\rho>1$, although the system has symmetric, i.\,e.,
periodic boundary conditions. However, this sign change of the Casimir
force at $\rho=1$ does not contradict the predictions of Bachas \cite{Bachas07},
as he assumed an infinite system in parallel direction, i.\,e., $\rho=0$.
The scaling function $\Xc(x,1)$ has negative slope $\Xc'(0,1)=-\Theta'(0,1)/d\nu$
at $x=0$. This behavior is in perfect agreement with Eq.~(\ref{eq:FC_von_uex}),
as the excess internal energy $u_{\mathrm{ex}}(\Tc,L,L)$ is negative
for our model. Furthermore, $\Xc(x,1)$ has a second zero at $x=-2.25(5)$
where $u(T,L,L)=u_{\infty}(T)$ holds. Fig.~\ref{fig:CasimirPlots}d
shows results from both the calculation using Eqs.~(\ref{eq:u}-\ref{eq:FC_integral})
(open symbols) as well as Eq.~(\ref{eq:FC_von_uex}) (filled symbols),
where the latter have a much better statistics, as no numerical differentiation
and integration is necessary.

Finally, in Fig.~\ref{fig:theta_p} we depict the Casimir scaling
function for values of $\rho$ larger than one. Now we are in rod
geometry and use the appropriate scaling variable $\Lp$ instead of
$\Ls$. Due to this rescaling, the scaling function $\Xc_{\parallel}(x_{\parallel},\rho)$
converges to a finite limit $\Xc_{\parallel}(x_{\parallel},\infty)$
which should only slightly deviate from curves for $\rho=8$, just
as in the inverse case $\rho=1/8$ (see Fig.~\ref{fig:FC1}). In
this regime the Casimir force is always positive, leading to a repulsion
of the opposite surfaces. Note that for $\rho=8$ we increased the
thickness difference for the calculation of the derivative in Eq.~(\ref{eq:FI})
to $\Ls'=\Ls\pm4$, as, e.\,g., $\Ls=256$ for $\Lp=32$.

\subsection{Excess free energy \label{sub:Excess-free-energy}}

\begin{figure}
\begin{centering}
\includegraphics[scale=0.6]{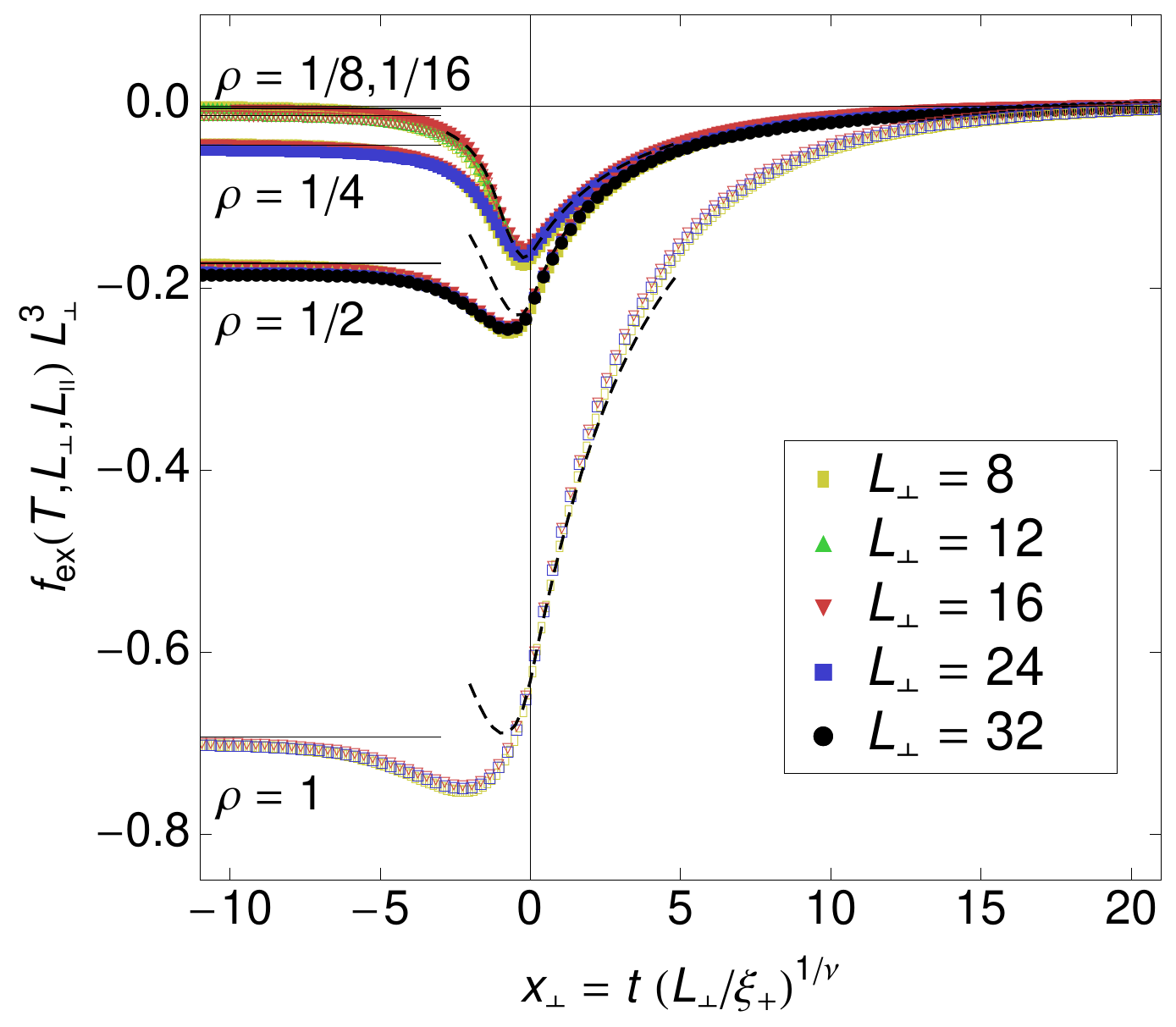}
\par\end{centering}

\caption{(Color online) Excess free energy scaling function $\Theta_{\perp}(x_{\perp},\rho)$
for several aspect ratios $\rho$. The dashed lines are the predictions
of Dohm \cite{Dohm09} for $\rho=1/4,1/2,1$, while the solid lines
are the limits for $x_{\perp}\to-\infty$, Eq.~(\ref{eq:Theta_s-inf-d}).
\label{fig:f_ex_3d}}

\end{figure}

The excess free energy is shown in Fig.~\ref{fig:f_ex_3d} for $\rho\leq1$.
An interesting feature of these curves is the non-vanishing limit
for $x_{\perp}\to-\infty$, which means that for fixed temperatures
$T<\Tc$ and $\Ls,\Lp\to\infty$ the total excess free energy $Vf_{\mathrm{ex}}$
approaches a finite value. This behavior is a direct consequence of
the broken symmetry in the ordered phase \cite{PrivmanFisher83}:
In this phase, which only exists in the thermodynamic limit below
$\Tc$, the Ising partition function is reduced by a factor of two,
as the system cannot reach the whole phase space anymore. This leads
to the term $-\ln2$ in the total excess free energy of a periodic
Ising system below $\Tc$, \begin{equation}
\Theta(-\infty,\rho)=-\ln2,\label{eq:Theta-inf}\end{equation}
\emph{independent} of shape and dimensionality. Note that, e.\,g.,
for the $q$-state Potts model this argument directly generalizes
to $\Theta(-\infty,\rho)=-\ln q$. Using Eq.~(\ref{eq:Theta_s}),
we find \begin{equation}
\Theta_{\perp}(-\infty,\rho)=-\rho^{d-1}\ln2,\label{eq:Theta_s-inf-d}\end{equation}
this limit is shown as thin solid lines in Fig.~\ref{fig:f_ex_3d}.
The results are compared to the field theoretical predictions of Dohm
\cite{Dohm09}, we find a satisfactory agreement for positive and
also for slightly negative values of $x_{\perp}$. Furthermore, our
value $\Delta(1)=-0.63(1)$ for the cube is compatible with the value
$-0.657(30)$ obtained by Mon \cite{Mon85}.$ $

\begin{figure}[b]
\begin{centering}
\includegraphics[scale=0.6]{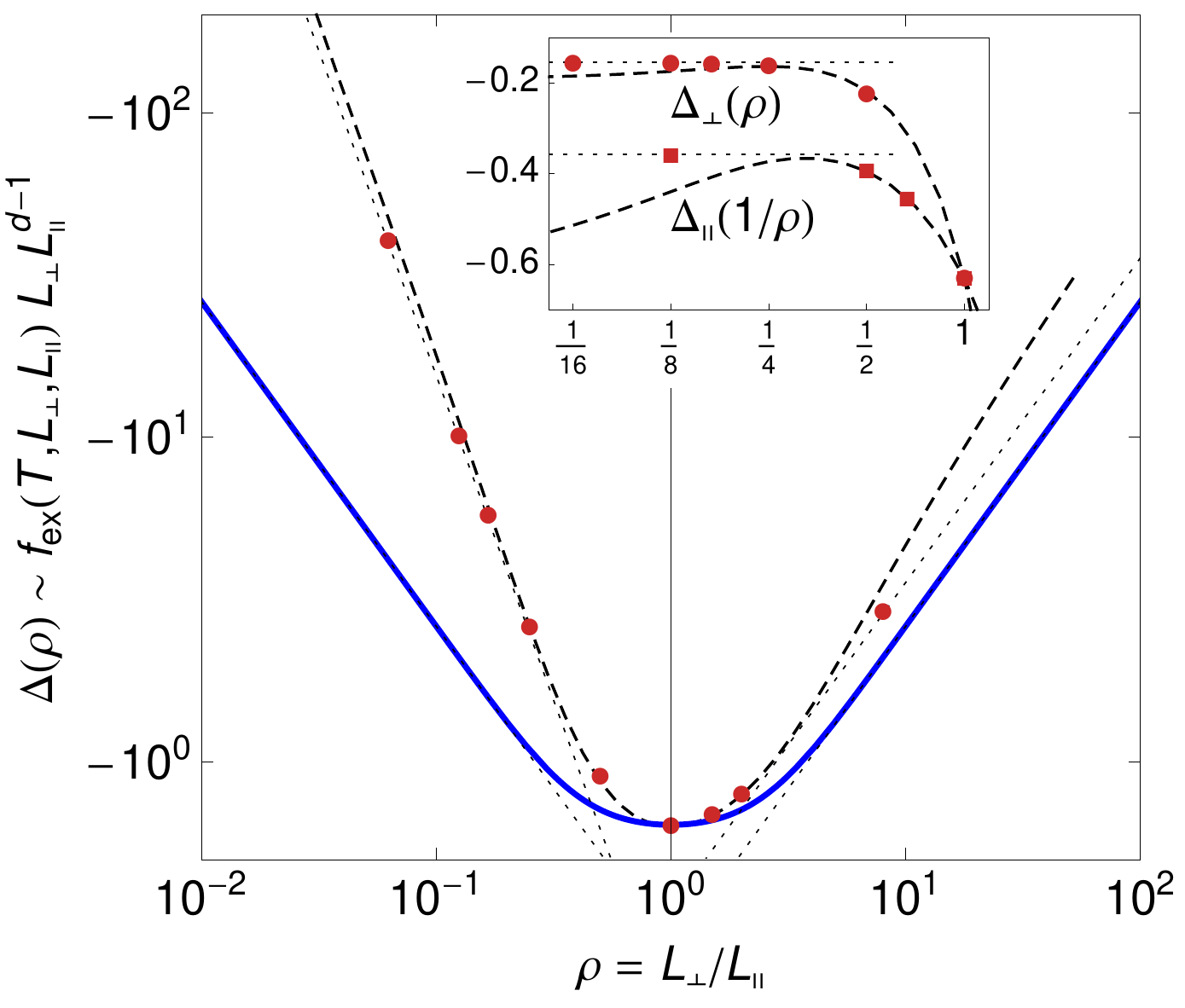}
\par\end{centering}

\caption{(Color online) Generalized Casimir amplitude $\Dc(\rho)=\Theta(0,\rho)$,
Eq.~(\ref{eq:Delta}), of the Ising universality class in $d=3$
(red circles, see also Tab.~\ref{tab:Casimir-amplitudes}) and in
$d=2$ (Eq.~(\ref{eq:Theta_p_2d_x0}), blue solid line). The dashed
line is the prediction of Dohm \cite{Dohm09}, while the dotted lines
show the asymptotes. The inset depicts $\Delta_{\perp}(\rho)$ (circles)
and $\Delta_{\parallel}(1/\rho)$ (squares). \label{fig:f_ex_0_3d}}

\end{figure}

The generalized Casimir amplitude at criticality, $\Dc(\rho)=\Theta(0,\rho)$
(Eq.~(\ref{eq:Delta})), is listed in Tab.~\ref{tab:Casimir-amplitudes}
for several values of $\rho$ and is depicted in Fig.~\ref{fig:f_ex_0_3d},
together with the predictions of Dohm \cite{Dohm09} (dashed line)
as well as the asymptotes (dotted lines). The inset shows $\Delta_{\perp}(\rho)$
(circles) and $\Delta_{\parallel}(1/\rho)$ (squares), showing good
agreement with these predictions for $1/4\lesssim\rho\lesssim3$.
\begin{table}[b]
\caption{Monte Carlo results for the Casimir amplitudes $\Delta(\rho)$, $\Delta_{\mu}(\rho)$
and $\Xc_{\mu}(0,\rho)$, with $\mu=\perp$ for $\rho\leq1$ and $\mu=\parallel$
for $\rho\geq1$. Note that the critical Casimir force changes sign
at $\rho=1$.\label{tab:Casimir-amplitudes}}

\centering{}\begin{tabular}{>{\centering}p{0.22\columnwidth}>{\centering}p{0.22\columnwidth}>{\centering}p{0.22\columnwidth}>{\centering}p{0.22\columnwidth}}
\hline 
$\rho$ & $\Delta(\rho)$ & $\Delta_{\mu}(\rho)$ & $\Xc_{\mu}(0,\rho)$\tabularnewline
\hline
$0$ & $-\infty$ & $-0.155(3)$ & $-0.310(6)$\tabularnewline
$1/16$ & $-39.8(8)$ & $-0.155(3)$ & $-0.310(6)$\tabularnewline
$1/8$ & $-9.9(2)$ & $-0.155(3)$ & $-0.310(6)$\tabularnewline
$1/6$ & $-5.7(1)$ & $-0.157(3)$ & $-0.30(1)$\tabularnewline
$1/4$ & $-2.60(5)$ & $-0.161(3)$ & $-0.290(5)$\tabularnewline
$1/2$ & $-0.89(2)$ & $-0.223(4)$ & $-0.22(1)$\tabularnewline
$1$ & $-0.63(1)$ & $-0.63(1)$ & $0.000(5)$\tabularnewline
$3/2$ & $-0.68(3)$ & $-0.45(2)$ & $0.17(1)$\tabularnewline
$2$ & $-0.78(3)$ & $-0.39(2)$ & $0.25(1)$\tabularnewline
$8$ & $-2.86(5)$ & $-0.357(8)$ & $0.36(1)$\tabularnewline
$\infty$ & $-\infty$ & $-0.36(1)$ & $0.36(1)$\tabularnewline
\hline
\end{tabular}
\end{table}

\section{Exact results in two dimensions\label{sec:2d-system}}

The scaling function $\Theta_{\perp}$ of the excess free energy in
$d=2$ is calculated exactly based on the work of Ferdinand and Fisher
\cite{FerdinandFisher69} (Note that the term $\xi S_{1}(n)\tau^{2}/2$
is missing in Eq.~(3.36) of this work). Our scaling variables differ
from theirs, we use $x_{\perp}=t(\Ls/\xip)$ and $\rho=\Ls/\Lp$,
while they used $\tau=x_{\perp}/2$ and $\xi=1/\rho$ as temperature
and aspect-ratio variables. 

We start from the partition function of the $\Ls\times\Lp$ isotropic
Ising model on a torus \cite{Kaufman49}, \begin{subequations} \begin{eqnarray}
Z(T,\Ls,\Lp) & = & \frac{1}{2}(2\sinh2K)^{\frac{1}{2}\Ls\Lp}\times\nonumber \\
 &  & \times\left(Z_{1}^{+}+Z_{1}^{-}+Z_{0}^{+}\pm Z_{0}^{-}\right),\label{eq:Z_Ising}\end{eqnarray}
with $+$ above and $-$ below $\Tc$, the four partial sums\begin{equation}
Z_{\delta}^{\pm}=\prod_{n=0}^{\Ls-1}\left(e^{\frac{1}{2}\Lp\gamma_{2n+\delta}}\pm e^{\frac{1}{2}\Lp\gamma_{2n+\delta}}\right),\label{eq:Z_partial}\end{equation}
 and $\cosh\gamma_{l}=\cosh2K\coth2K-\cos(l\pi/\Ls)$. \end{subequations}

For the bulk free energy density of the $d=2$ Ising model we using
\emph{Mathematica} \cite{MMA7} derived a nice closed-form expression
not present in the literature yet, namely \begin{equation}
f_{\infty}=-\ln(2\cosh2K)+\frac{k^{2}}{16}\,{_{4}F_{3}}\!\left(\left.{1,1,{\textstyle \frac{3}{2}},{\textstyle \frac{3}{2}}\atop 2,2,2}\right|k^{2}\right)\label{eq:f_2d}\end{equation}
with $k=2\tanh2K/\cosh2K$ and the generalized hypergeometric function
${_{4}F_{3}}(\cdot)$ \cite{MMA7}.

\begin{figure}
\begin{centering}
\includegraphics[scale=0.6]{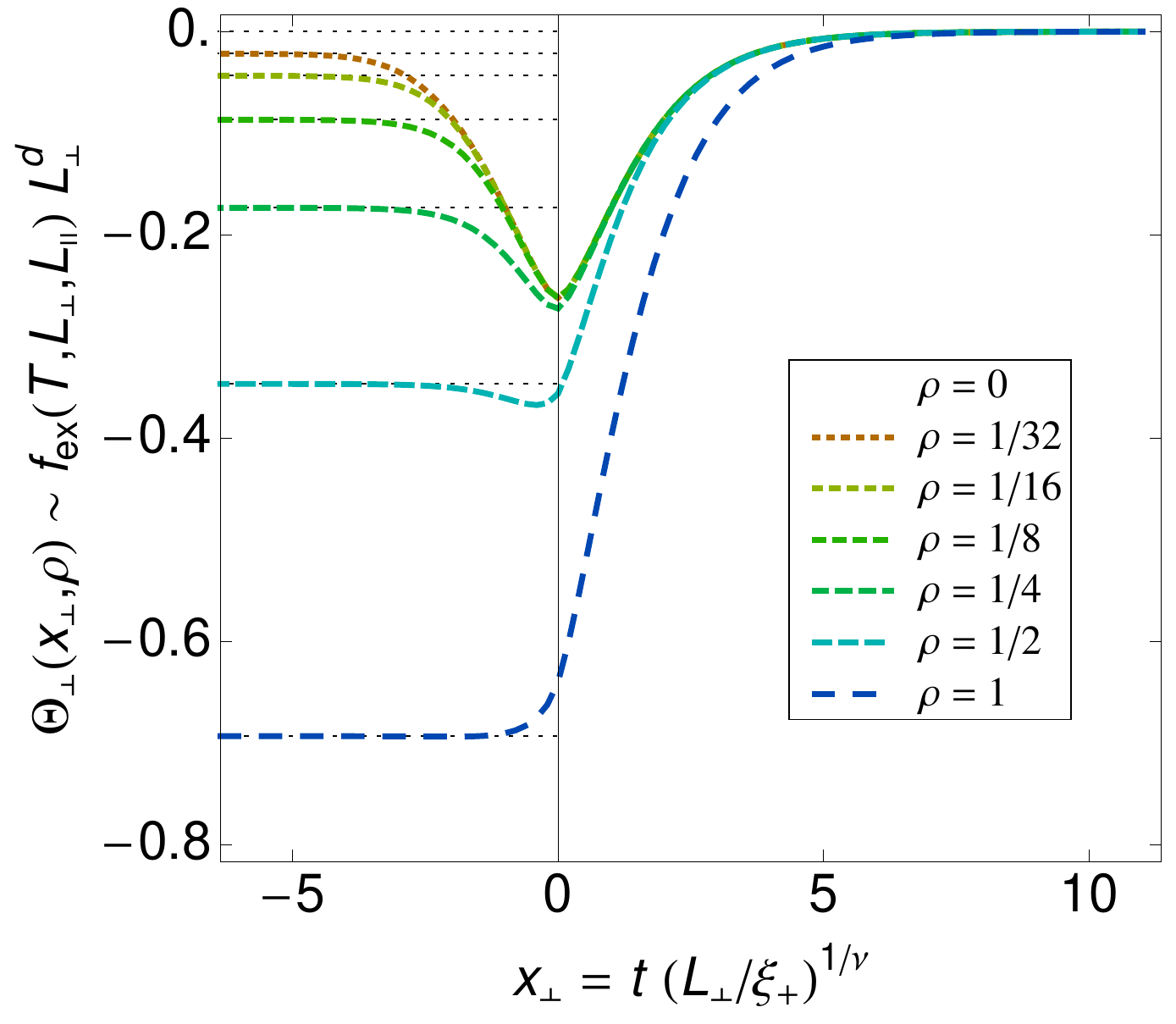}
\par\end{centering}

\caption{(Color online) Excess free energy scaling function of the $d=2$ Ising
model for several aspect ratios $\rho\le1$. The scaling functions
for $\rho\ge1$ can be calculated using Eq.~(\ref{eq:Theta_s_2d_symm}).
\label{fig:Theta_2d}}

\end{figure}
\begin{figure}
\begin{centering}
\includegraphics[scale=0.6]{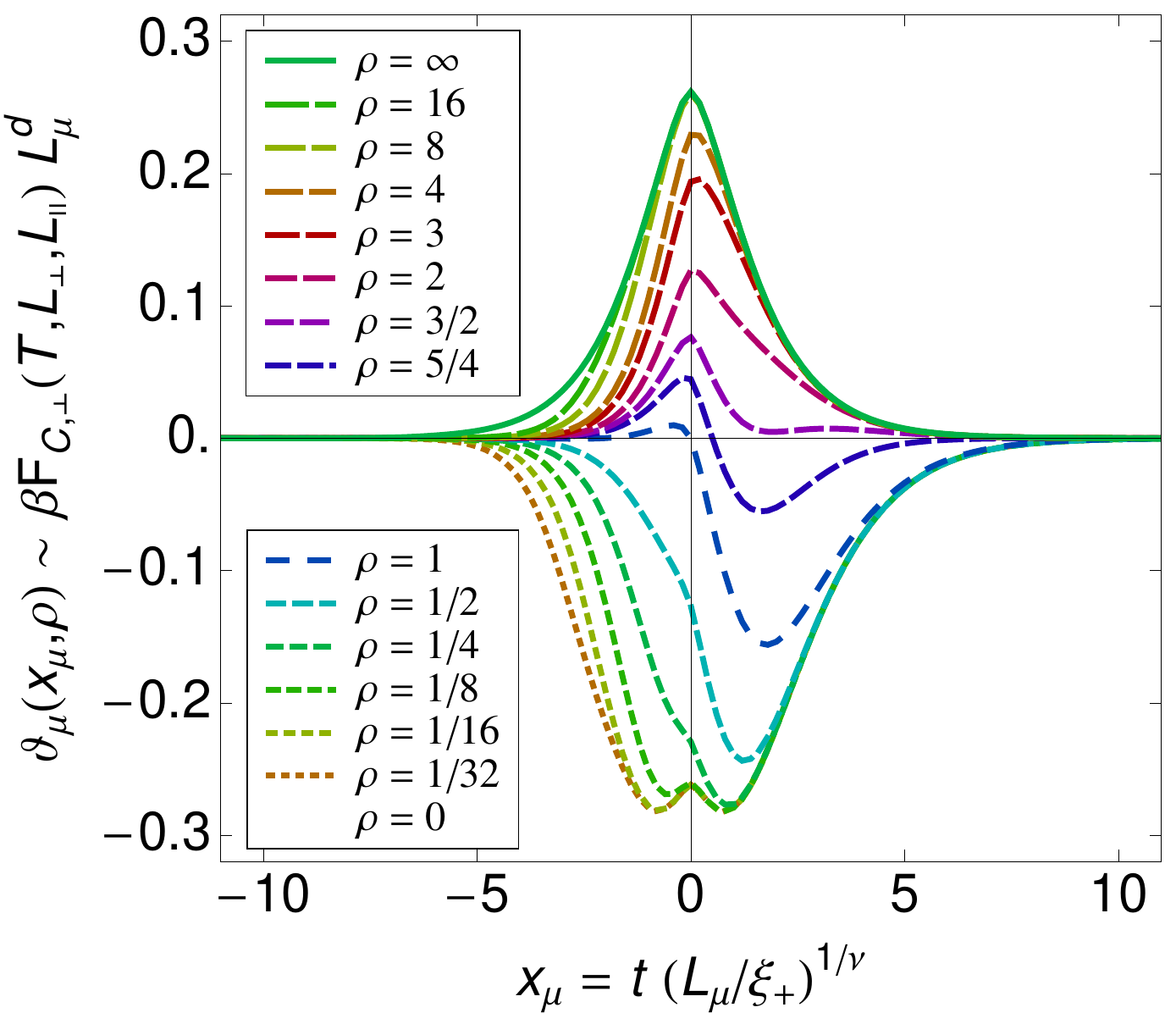}
\par\end{centering}

\caption{(Color online) Casimir force scaling function $\Xc_{\mu}(x_{\mu},\rho)$
of the $d=2$ Ising model for several aspect ratios $\rho$. Shown
is $\Xc_{\perp}(x_{\perp},\rho)$ for $\rho\le1$ and $\Xc_{\parallel}(x_{\parallel},\rho)$
for $\rho\ge1$. Note that $\Xc_{\perp}(x_{\perp},\rho)=\rho^{2}\Xc_{\parallel}(x_{\parallel},\rho)$.
\label{fig:theta_2d}}

\end{figure}
After some algebra, the scaling function $\Theta_{\perp}$ for arbitrary
$x_{\perp}$ and $\rho$ can be written as \begin{subequations}\label{eq:Theta_p_2d}\begin{equation}
\Theta_{\perp}(x_{\perp},\rho)=-\rho\ln\!\left(\frac{P_{\nicefrac{1}{2}}^{+}+P_{\nicefrac{1}{2}}^{-}}{2e^{-I_{+}/\rho}}+\frac{P_{\vphantom{\nicefrac{1}{2}}0}^{+}\pm P_{0}^{-}}{2e^{-I_{-}/\rho}}\right)\label{eq:Theta_p_2d_}\end{equation}
with\begin{equation}
P_{\delta}^{\pm}(x_{\perp},\rho)=\prod_{n=-\infty}^{\infty}\left(1\pm e^{-\sqrt{x_{\perp}^{2}+4\pi^{2}(n-\delta)^{2}}/\rho}\right)\label{eq:Ppm}\end{equation}
and\begin{equation}
I_{\pm}(x_{\perp})=\int_{-\infty}^{\infty}\mathrm{d}\omega\ln\!\left(1\pm e^{-\sqrt{x_{\perp}^{2}+4\pi^{2}\omega^{2}}}\right).\label{eq:Ipm}\end{equation}
\end{subequations}Note that\begin{equation}
I_{\pm}(x_{\perp})=\lim_{r\to\infty}\frac{1}{r}\ln P_{\delta}^{\pm}(rx_{\perp},r)\label{eq:IpmLim}\end{equation}
independent of $\delta$. As the $2d$ system is invariant under exchange
of the directions $\perp$ and $\parallel$, \C{i.\,e., $\rho\to1/\rho$
and $x_{\perp}\to x_{\parallel}=x_{\perp}/\rho$,} \begin{equation}
\Theta(x,\rho)=\Theta(x,1/\rho),\label{eq:Theta_2d_symm}\end{equation}
which using Eq.~(\ref{eq:Theta_s}) gives\begin{equation}
\Theta_{\perp}(x_{\perp},\rho)/\rho=\rho\Theta_{\perp}(x_{\perp}/\rho,1/\rho),\label{eq:Theta_s_2d_symm}\end{equation}
we can derive the identities\begin{subequations}\begin{eqnarray}
\frac{P_{\nicefrac{1}{2}}^{+}(x_{\perp},\rho)}{P_{\nicefrac{1}{2}}^{+}(x_{\perp}/\rho,1/\rho)} & = & \frac{e^{\rho I_{+}(x_{\perp}/\rho)}}{e^{I_{+}(x_{\perp})/\rho}},\\
\frac{P_{\nicefrac{1}{2}}^{-}(x_{\perp},\rho)}{P_{0}^{+}(x_{\perp}/\rho,1/\rho)} & = & \frac{e^{\rho I_{-}(x_{\perp}/\rho)}}{e^{I_{+}(x_{\perp})/\rho}},\\
\frac{P_{0}^{-}(x_{\perp},\rho)}{P_{0}^{-}(x_{\perp}/\rho,1/\rho)} & = & \frac{e^{\rho I_{-}(x_{\perp}/\rho)}}{e^{I_{-}(x_{\perp})/\rho}},\end{eqnarray}
\end{subequations}which are a generalization of Jacobi's imaginary
transformations for elliptic $\Xc$ functions \cite{WhittakerWatson90}. 

The resulting excess free energy scaling function $\Theta_{\perp}(x_{\perp},\rho)$
is depicted in Fig.~\ref{fig:Theta_2d}, showing a similar behavior
as in the three-dimensional case. For $x_{\perp}\to-\infty$ Eq.~(\ref{eq:Theta_p_2d})
simplifies to \begin{equation}
\Theta_{\perp}(-\infty,\rho)=-\rho\ln2,\label{eq:Theta_p-inf}\end{equation}
as explained in Sec.~\ref{sub:Excess-free-energy}.

\begin{table}[b]
\caption{Signs of the terms $P_{\delta}^{\pm}(x_{\perp},\rho)$ in Eq.~(\ref{eq:Theta_p_2d_})
for different boundary conditions. \label{tab:signs}}

\centering{}\begin{tabular}{>{\centering}p{0.22\columnwidth}>{\centering}p{0.22\columnwidth}>{\centering}p{0.1\columnwidth}>{\centering}p{0.1\columnwidth}>{\centering}p{0.1\columnwidth}>{\centering}p{0.1\columnwidth}}
\hline 
$\mathrm{BC}_{\perp}$ & $\mathrm{BC}_{\parallel}$ & $P_{\nicefrac{1}{2}}^{+}$ & $P_{\nicefrac{1}{2}}^{-}$ & $P_{0}^{+}$ & $P_{0}^{-}$\tabularnewline
\hline
periodic & periodic & $+$ & $+$ & $+$ & $-$\tabularnewline
periodic & antiperiodic & $+$ & $+$ & $-$ & $+$\tabularnewline
antiperiodic & periodic & $+$ & $-$ & $+$ & $+$\tabularnewline
antiperiodic & antiperiodic & $-$ & $+$ & $+$ & $+$\tabularnewline
\hline
\end{tabular}
\end{table}
From Eq.~(\ref{eq:Theta_p_2d}) we directly obtain values of the
scaling function at the critical point $x_{\perp}=0$, as \begin{subequations}
\begin{equation}
I_{+}(0)=\frac{\pi}{12},\qquad I_{-}(0)=-\frac{\pi}{6},\label{eq:I(0)}\end{equation}
and\begin{equation}
P_{\nicefrac{1}{2}}^{\pm}(0,\rho)=(\mp q;q^{2})_{\infty}^{2},\quad P_{0}^{\pm}(0,\rho)=\frac{1}{2}(\mp1;q^{2})_{\infty}^{2}\label{eq:P(0,rho)}\end{equation}
\end{subequations}with $q=e^{-\pi/\rho}$ and the $q$-Pochhammer
symbol \cite{MMA7} $(a;q)_{\infty}$, leading to\begin{eqnarray}
\Theta_{\perp}(0,\rho) & = & -\rho\ln\!\left(\frac{(-q;q^{2})_{\infty}^{2}+(q;q^{2})_{\infty}^{2}}{2q^{1/12}}+\frac{(-1;q^{2})_{\infty}^{2}}{4q^{-1/6}}\right)\nonumber \\
 & = & -\rho\ln\frac{\Xc_{2}(0,q)+\Xc_{3}(0,q)+\Xc_{4}(0,q)}{(4\Xc_{2}(0,q)\Xc_{3}(0,q)\Xc_{4}(0,q))^{1/3}}\label{eq:Theta_p_2d_x0}\end{eqnarray}
 after expressing the $q$-Pochhammer symbols in terms of elliptic
$\Xc$ functions. This result was already given by Ferdinand and Fisher
\cite{FerdinandFisher69} (Eq.~(3.37)). The resulting Casimir amplitude
$\Dc(\rho)$ is shown as blue solid line in Fig.~\ref{fig:f_ex_0_3d}.

From the exact solution Eq.~(\ref{eq:Theta_p_2d}) we calculated
the Casimir force scaling function by numerical differentiation using
the scaling relation Eq.~(\ref{eq:ScalingIdentity_s}), as an analytic
derivation would be too lengthy for arbitrary $\rho$. The results
are shown in Fig.~\ref{fig:theta_2d}, for $\rho\leq1$ we show $\Xc_{\perp}(x_{\perp},\rho)$,
while for $\rho\geq1$ we show $\Xc_{\parallel}(x_{\parallel},\rho).$
Clearly the Casimir force changes sign from negative to positive values
with increasing aspect ratio $\rho$, as in the three-dimensional
case.

Finally we give expressions for the limits $\rho\to0$ and $\rho\to\infty$.
In film geometry, $\rho\to0$, Eq.~(\ref{eq:Theta_p_2d}) reduces
to the simple result \begin{eqnarray}
\Theta_{\perp}(x_{\perp},0) & = & -I_{+}(x_{\perp})\nonumber \\
 & = & -\frac{1}{\pi}\int_{0}^{\infty}\mathrm{d}\omega\ln\left(1+e^{-\sqrt{x_{\perp}^{2}+\omega^{2}}}\right),\label{eq:Theta_s_2d_rho0}\end{eqnarray}
yielding the already exactly known Casimir force scaling function
\cite{RudnickZandiShackellAbraham10} \begin{equation}
\Xc_{\perp}(x_{\perp},0)=-\frac{1}{\pi}\int_{0}^{\infty}\mathrm{d}\omega\frac{\sqrt{x_{\perp}^{2}+\omega^{2}}}{1+e^{\sqrt{x_{\perp}^{2}+\omega^{2}}}}.\label{eq:theta_s_2d_rho0}\end{equation}

In the opposite limit $\rho\to\infty$ we have \begin{equation}
\Theta_{\parallel}(x_{\parallel},\infty)=-\vartheta_{\parallel}(x_{\parallel},\infty)=-I_{+}(x_{\parallel})\label{eq:theta_p_2d_rhoinf}\end{equation}
using Eq.~(\ref{eq:ScalingIdentity_inf}). For both $\rho\to0$ and
$\rho\to\infty$ we have the symmetries $\Theta_{\perp}(x_{\perp},\rho)=\Theta_{\perp}(-x_{\perp},\rho)$
and $\Xc_{\perp}(x_{\perp},\rho)=\Xc_{\perp}(-x_{\perp},\rho)$. Note
that all scaling predictions from the previous sections have been
verified in the $d=2$ Ising case. Finally, we remark that these calculations
can be easily extended to mixed periodic-antiperiodic boundary conditions
by modifying the prefactors of the four terms $P_{\delta}^{\pm}(x_{\perp},\rho)$
in Eq.~(\ref{eq:Theta_p_2d_}) according to Tab.~\ref{tab:signs}.

\section{Summary\label{sec:Summary}}

In this work we calculated the universal excess free energy and Casimir
force scaling functions, $\Theta(x,\rho)$ and $\vartheta(x,\rho)$,
of the three- and two-dimensional Ising model with arbitrary aspect
ratio $\rho$ and periodic boundary conditions in all directions.
In $d=3$ we used Monte Carlo simulations based on the method by Hucht
\cite{Hucht07a}, while in $d=2$ we derived an analytic expression,
Eq.~(\ref{eq:Theta_p_2d}), for the excess free energy scaling function
$\Theta(x,\rho)$. Furthermore, we derived several new scaling identities
for the scaling functions: We showed that the Casimir force scaling
function $\vartheta_{\perp}(x_{\perp},0)$ in the film limit has a
singularity of order $(x_{\perp}-x_{\perp}^{*})^{1-\alpha^{*}}$ at
the point $x_{\perp}^{*}<0$ where the $d{-}1$-dimensional system
has a phase transition (Eq.~(\ref{eq:theta(x*)})), where $\alpha^{*}$
denotes the specific heat exponent of the $d{-}1$-dimensional system.
In our case $\alpha^{*}=0$ and the singularity is logarithmic as
shown in Figs.~\ref{fig:Fint1} and \ref{fig:FC1}. At finite values
of $\rho\gtrsim1/4$ our results are compared to field-theoretical
results of Dohm, and we find good agreement in the regime $x\gtrsim-2$
where his theory is expected to be valid \cite{Dohm09}. For the cube
with $\rho=1$ we observed another interesting result, here the Casimir
force vanishes exactly at the critical point, $\vartheta(0,1)=0$.
In Appendix~\ref{sec:Stationarity-of-Theta} this property is shown
to hold for all systems that are invariant under permutation of the
directions, and is not restricted to periodic systems. The vanishing
Casimir force could serve as a stability/instability criterion with
respect to $\rho$: If we assume that the system can change the lengths
$L_{\mu}$ at constant volume, we see that the cube with $\rho=1$
and periodic boundary conditions is unstable under variation of $\rho$
at $x=0$, as $\rho<1$ tends to $\rho\to0$ and $\rho>1$ tends to
$\rho\to\infty$. Note that this behavior would reverse for antiperiodic
boundary conditions, then the cube would be stable at $x=0$ and the
equilibrium shape would even be temperature dependent, as the zero
of $\vartheta(x,\rho)$ varies with $x$, see Fig.~\ref{fig:theta_2d}.
For $\rho>1$ the Casimir force is positive and converges against
the negative excess free energy, $\Xc_{\parallel}(x_{\parallel},\infty)=-\Theta_{\parallel}(x_{\parallel},\infty)$,
Eq.~(\ref{eq:ScalingIdentity_inf}). 

The excess free energy below $\Tc$ is $f_{\mathrm{ex}}\sim-V^{-1}\ln2$
in periodic Ising systems \cite{PrivmanFisher83} independent of system
shape (Eq.~(\ref{eq:Theta-inf})), leading to a finite $\rho$-dependent
limit of $\Theta_{\perp}(-\infty,\rho)$, Eq.~(\ref{eq:Theta_s-inf-d}). 

Finally, the universal scaling function $\Theta_{\perp}(x_{\perp},\rho)$
is calculated exactly in $d=2$, and the results are found to be in
qualitative agreement with the results for $d=3$. The most important
difference between these two cases is the fact that the $2d$ system
has several symmetries not present in the $3d$ system, i.\,e. $(x_{\perp},\rho)\leftrightarrow(x_{\parallel},1/\rho)$,
$(x,0)\leftrightarrow(-x,0)$, and $(x,\infty)\leftrightarrow(-x,\infty)$. 
\begin{acknowledgments}
One of the authors (AH) would like to thank Martin Hasenbusch for
very useful discussions.
\end{acknowledgments}
\appendix

\section{Stationarity of $\Theta(x,\rho)$ at $\rho=1$ \label{sec:Stationarity-of-Theta}}

\begin{figure}
\begin{centering}
\includegraphics[scale=0.6]{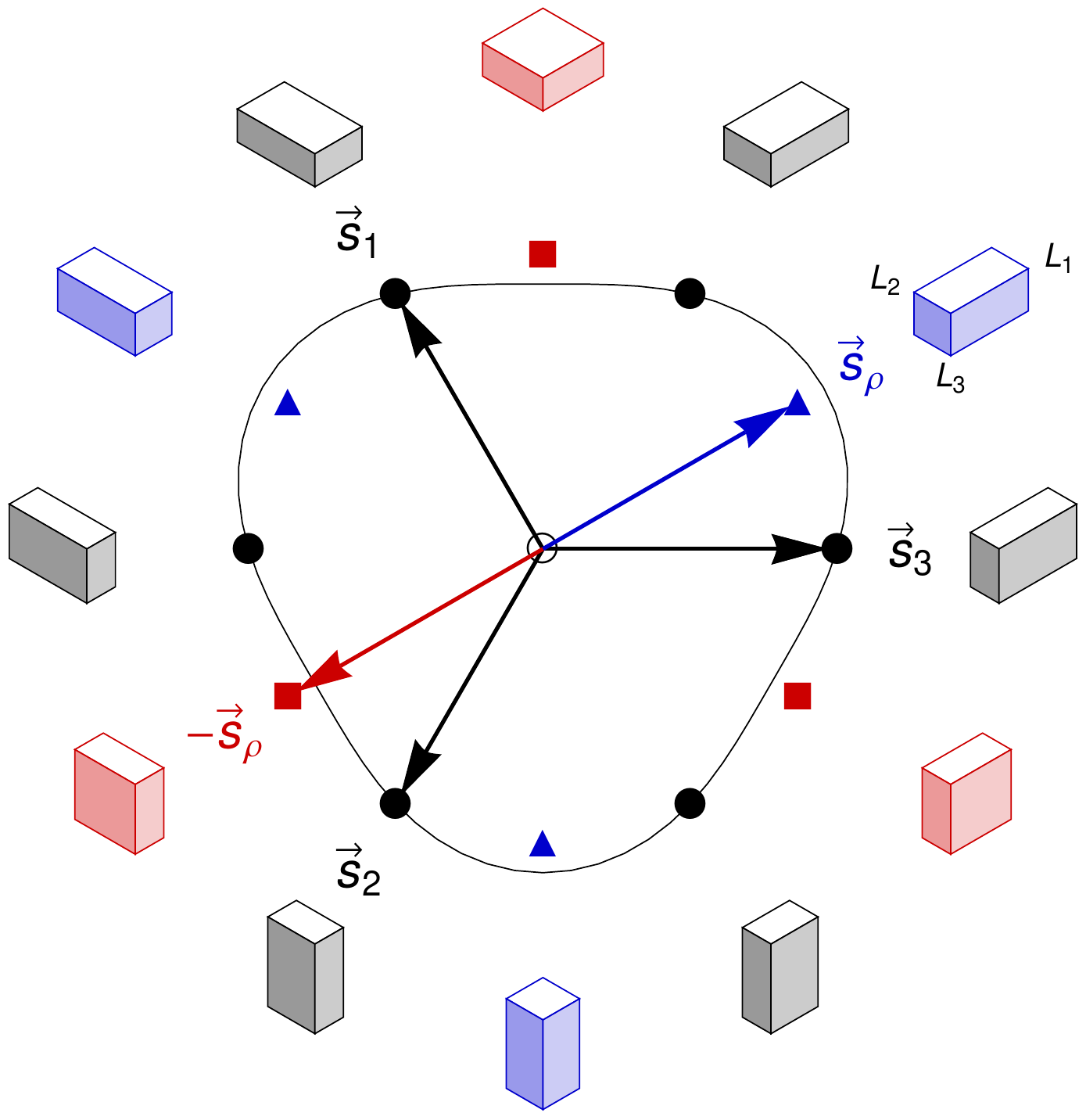}
\par\end{centering}

\caption{(Color online) The $d{-}1$-dimensional plane $\mathcal{B}$ of constant
volume $L^{d}$ for $d=3$, viewed from the normal direction $(1,1,1)$.
The origin at the center ($\circ$) is the cube with $\vec{\B}=\vec{0}$,
while the filled symbols are deformed systems as indicated by the
pictures: The black points mark the directions $\pm\vec{s}_{\mu}$
with constant $L_{\mu}$ symmetric under permutation $\mathcal{P}$,
Eq.~(\ref{eq:permutation}). The blue arrow $\vec{s}_{\rho}$ ($\rho>1$)
and the red arrow $-\vec{s}_{\rho}$ ($\rho<1$) mark the direction
of the shape variation in terms of $\rho$ used in this work. The
black curve sketches a line of constant $\Theta(x,\vec{\B})$. Note
that $\Theta(x,\vec{s}_{\rho})\neq\Theta(x,-\vec{s}_{\rho})$, as
the shape and thus $\Theta$ is not symmetric under the transformation
$\rho\to1/\rho$, see also Fig.~\ref{fig:f_ex_0_3d}. \label{fig:Shape-Plane}}

\end{figure}
The stationarity of the excess free energy scaling function $\Theta(x,\rho)$
at $\rho=1$ can be derived for isotropic systems with arbitrary symmetric
boundary conditions and in arbitrary dimensions $d$: We allow arbitrary
shape changes of $f_{\mathrm{ex}}(T,L_{1},\ldots,L_{d})$ and write
$L_{\mu}=e^{\B_{\mu}}L$, so that Eq.~(\ref{eq:FSS_fex}) now reads

\begin{equation}
f_{\mathrm{ex}}(T,e^{\B_{1}}L,\ldots,e^{\B_{d}}L)\sim L^{-d}\Theta(x,\vec{\B}),\label{eq:f_ex_d}\end{equation}
under the condition \begin{equation}
\sum_{\mu=1}^{d}\B_{\mu}=0\label{eq:b.eq.0}\end{equation}
defining the plane $\mathcal{B}$ with constant volume $L^{d}$. The
symmetry under permutation of the $d$ lattice axes implies \begin{equation}
\Theta(x,\vec{\B})=\Theta(x,\mathcal{P}(\vec{\B}))\label{eq:permutation}\end{equation}
with permutation operator $\mathcal{P}$. This symmetry holds if the
boundary conditions in all directions are equal. Without loss of generality
we now assume $d=3$, $\B_{1}=0$ and vary the shape of the system
along directions $2$ and $3$, i.\,e., $\B_{2}=-\B_{3}$, so that
$\Theta(x,\epsilon\,\vec{s}_{1})=\Theta(x,-\epsilon\,\vec{s}_{1})$
with $\vec{s}_{1}=(0,1,-1)$ and real $\epsilon$. Hence $\Theta(x,\epsilon\,\vec{s}_{1})$
is an even function of $\epsilon$ and thus the directional derivative
along $\vec{s}_{1}$ at the origin vanishes, \begin{equation}
\left.\frac{\partial}{\partial\epsilon}\Theta(x,\epsilon\,\vec{s}_{1})\right|_{\epsilon=0}=0.\label{eq:directional_derivative}\end{equation}
The same argument holds for the symmetric directions $\vec{s}_{2}=(-1,0,1)$
and $\vec{s}_{3}=(1,-1,0)$ (see Fig.~\ref{fig:Shape-Plane}). As
the $d(d-1)/2$ vectors $\vec{s}_{\mu}$ form an (over)complete base
in the $d{-}1$-dimensional plane $\mathcal{B}$, and all directional
derivatives vanish at the origin $\vec{\B}=\vec{0}$, we conclude
that Eq.~(\ref{eq:directional_derivative}) holds for all directions
$\vec{s}\in\mathcal{B}$. Hence Eq.~(\ref{eq:directional_derivative})
also holds for the special case $\vec{s}_{\rho}=(2/3,-1/3,-1/3)$
which is the direction of the shape variation used in this work (with
$\rho=e^{\epsilon}$), if we set $L_{1}=\Ls$ and $L_{2}=L_{3}=\Lp$.
From this we conclude that Eq.~(\ref{eq:FSS_fex}) satisfies \begin{equation}
\left.\frac{\partial}{\partial\rho}\Theta(x,\rho)\right|_{\rho=1}=0.\label{eq:dTheta_drho}\end{equation}
Note that these arguments can be extended to weakly anisotropic systems,
while less is known in the strongly anisotropic case \cite{Hucht02a,BurgsmuellerDiehlShpot10}.

\section{Proof of $\boldsymbol{\Xc(0,1)=0}$ in the large-$\boldsymbol{n}$
limit}

In this appendix we show for the large-$n$ limit \cite{Dohm09} that
the finite-size scaling function of the thermodynamic Casimir force
vanishes at bulk criticality in the case of a cubic system geometry
$\rho=1$. To this end we start from the scaling function of the singular
free energy per volume given by Dohm (Eq.~(3) in \cite{Dohm09}),
together with the self-consistency equation for the parameter $P(x_{\perp},\rho)$
at $x_{\perp}=0$,\begin{equation}
P(0,\rho)=-4\pi\mathcal{G}_{1}(P(0,\rho)^{2},\rho),\label{eq:Pappendix}\end{equation}
and the functions $\mathcal{G}_{j}(P^{2},\rho)$ (Eq.~(4) in \cite{Dohm09}).
Introducing the parameter $\hat{P}(\rho)\equiv\rho^{\mu}P(0,\rho)$
and furthermore the integration variable $\hat{z}=\rho^{\delta}z$
in the integral $\mathcal{G}_{0}(P^{2},\rho)$, the value of the excess
free energy scaling function $\Theta_{\perp}$ (see Eq.~(\ref{eq:FSS_fex_s}))
at bulk criticality can be cast in the form\begin{equation}
\Theta_{\perp}(0,\rho)=\Delta_{\perp}(\rho)=\rho^{2}\Delta(\rho)\label{eq:Thetantoinfappendix}\end{equation}
upon setting $\mu=-2/3$ and $\delta=4/3$, where $\Delta(\rho)$
is given by\begin{eqnarray}
\Delta(\rho) & = & -\frac{\hat{P}(\rho)^{3}}{12\pi}+\frac{1}{2}\int_{0}^{\infty}\frac{\mathrm{d}\hat{z}}{\hat{z}}\exp\!\left(-\frac{\hat{z}\hat{P}(\rho)^{2}}{4\pi^{2}}\right)\nonumber \\
 &  & \times\Bigl[\left(\frac{\pi}{\hat{z}}\right)^{3/2}-K(\rho^{-4/3}\hat{z})K(\rho^{2/3}\hat{z})^{2}\Bigr].\label{eq:hatDeltaappendix}\end{eqnarray}
According to Eq.~(\ref{eq:FC(Delta)}) one has\begin{equation}
\vartheta_{\perp}(x_{\perp}=0,\rho=1)=-\Delta'(1),\end{equation}
where the derivative of $\Delta(\rho)$ with respect to $\rho$ at
$\rho=1$ becomes\begin{equation}
\Delta'(1)=-\frac{\hat{P}(1)\hat{P}^{\prime}(1)}{4\pi}\left[\hat{P}(1)+4\pi\mathcal{G}_{1}(\hat{P}(1)^{2},1)\right].\end{equation}
Since $\hat{P}(1)=P(0,1)$ is the solution to Eq.~(\ref{eq:Pappendix})
at $\rho=1$, the expression in square brackets vanishes and thus
$\Delta'(1)=0$.

\bibliographystyle{apsrev4-1}
\bibliography{\string~/TeX/bib/Physik.bib}

\begin{thebibliography}{64}%
\makeatletter
\providecommand \@ifxundefined [1]{%
 \@ifx{#1\undefined}
}%
\providecommand \@ifnum [1]{%
 \ifnum #1\expandafter \@firstoftwo
 \else \expandafter \@secondoftwo
 \fi
}%
\providecommand \@ifx [1]{%
 \ifx #1\expandafter \@firstoftwo
 \else \expandafter \@secondoftwo
 \fi
}%
\providecommand \natexlab [1]{#1}%
\providecommand \enquote  [1]{``#1''}%
\providecommand \bibnamefont  [1]{#1}%
\providecommand \bibfnamefont [1]{#1}%
\providecommand \citenamefont [1]{#1}%
\providecommand \href@noop [0]{\@secondoftwo}%
\providecommand \href [0]{\begingroup \@sanitize@url \@href}%
\providecommand \@href[1]{\@@startlink{#1}\@@href}%
\providecommand \@@href[1]{\endgroup#1\@@endlink}%
\providecommand \@sanitize@url [0]{\catcode `\\12\catcode `\$12\catcode
  `\&12\catcode `\#12\catcode `\^12\catcode `\_12\catcode `\%12\relax}%
\providecommand \@@startlink[1]{}%
\providecommand \@@endlink[0]{}%
\providecommand \url  [0]{\begingroup\@sanitize@url \@url }%
\providecommand \@url [1]{\endgroup\@href {#1}{\urlprefix }}%
\providecommand \urlprefix  [0]{URL }%
\providecommand \Eprint [0]{\href }%
\@ifxundefined \urlstyle {%
  \providecommand \doi  [0]{\begingroup \@sanitize@url \@doi}%
  \providecommand \@doi [1]{\endgroup \@@startlink {\doibase
  #1}doi:\discretionary {}{}{}#1\@@endlink }%
}{%
  \providecommand \doi  [0]{doi:\discretionary{}{}{}\begingroup
  \urlstyle{rm}\Url }%
}%
\providecommand \doibase [0]{http://dx.doi.org/}%
\providecommand \Doi [0]{\begingroup \@sanitize@url \@Doi }%
\providecommand \@Doi  [1]{\endgroup\@@startlink{\doibase#1}\@@Doi}%
\providecommand \@@Doi [1]{#1\@@endlink}%
\providecommand \selectlanguage [0]{\@gobble}%
\providecommand \bibinfo  [0]{\@secondoftwo}%
\providecommand \bibfield  [0]{\@secondoftwo}%
\providecommand \translation [1]{[#1]}%
\providecommand \BibitemOpen [0]{}%
\providecommand \bibitemStop [0]{}%
\providecommand \bibitemNoStop [0]{.\EOS\space}%
\providecommand \EOS [0]{\spacefactor3000\relax}%
\providecommand \BibitemShut  [1]{\csname bibitem#1\endcsname}%
\bibitem [{\citenamefont {Casimir}(1948)}]{Casimir48}%
  \BibitemOpen
  \bibfield  {author} {\bibinfo {author} {\bibfnamefont {H.~B.~G.}\
  \bibnamefont {Casimir}},\ }\href@noop {} {\bibfield  {journal} {\bibinfo
  {journal} {Proc. K. Ned. Akad. Wet.},\ }\textbf {\bibinfo {volume} {51}},\
  \bibinfo {pages} {793} (\bibinfo {year} {1948})}\BibitemShut {NoStop}%
\bibitem [{\citenamefont {Lamoreaux}(1997)}]{Lamoreaux97}%
  \BibitemOpen
  \bibfield  {author} {\bibinfo {author} {\bibfnamefont {S.~K.}\ \bibnamefont
  {Lamoreaux}},\ }\Doi {10.1103/PhysRevLett.78.5} {\bibfield  {journal}
  {\bibinfo  {journal} {Phys. Rev. Lett.},\ }\textbf {\bibinfo {volume} {78}},\
  \bibinfo {pages} {5} (\bibinfo {year} {1997})},\ \bibinfo {note} {{Phys.}
  {Rev.} {Lett.}, {\bf 81}, 5475 (1998) (Erratum), arXiv:1007.4276}\BibitemShut
  {NoStop}%
\bibitem [{\citenamefont {Mohideen}\ and\ \citenamefont
  {Roy}(1998)}]{MohideenRoy98}%
  \BibitemOpen
  \bibfield  {author} {\bibinfo {author} {\bibfnamefont {U.}~\bibnamefont
  {Mohideen}}\ and\ \bibinfo {author} {\bibfnamefont {A.}~\bibnamefont {Roy}},\
  }\Doi {10.1103/PhysRevLett.81.4549} {\bibfield  {journal} {\bibinfo
  {journal} {Phys. Rev. Lett.},\ }\textbf {\bibinfo {volume} {81}},\ \bibinfo
  {pages} {4549} (\bibinfo {year} {1998})}\BibitemShut {NoStop}%
\bibitem [{\citenamefont {Fisher}\ and\ \citenamefont
  {de~Gennes}(1978)}]{FisherdeGennes78}%
  \BibitemOpen
  \bibfield  {author} {\bibinfo {author} {\bibfnamefont {M.~E.}\ \bibnamefont
  {Fisher}}\ and\ \bibinfo {author} {\bibfnamefont {P.-G.}\ \bibnamefont
  {de~Gennes}},\ }\href@noop {} {\bibfield  {journal} {\bibinfo  {journal} {C.
  R. Acad. Sci. Paris, Ser. B},\ }\textbf {\bibinfo {volume} {287}},\ \bibinfo
  {pages} {207} (\bibinfo {year} {1978})}\BibitemShut {NoStop}%
\bibitem [{\citenamefont {Gambassi}(2009)}]{Gambassi09a}%
  \BibitemOpen
  \bibfield  {author} {\bibinfo {author} {\bibfnamefont {A.}~\bibnamefont
  {Gambassi}},\ }\href {http://stacks.iop.org/1742-6596/161/i=1/a=012037}
  {\bibfield  {journal} {\bibinfo  {journal} {Journal of Physics: Conference
  Series},\ }\textbf {\bibinfo {volume} {161}},\ \bibinfo {pages} {012037}
  (\bibinfo {year} {2009})}\BibitemShut {NoStop}%
\bibitem [{\citenamefont {Garcia}\ and\ \citenamefont
  {Chan}(1999)}]{GarciaChan99}%
  \BibitemOpen
  \bibfield  {author} {\bibinfo {author} {\bibfnamefont {R.}~\bibnamefont
  {Garcia}}\ and\ \bibinfo {author} {\bibfnamefont {M.~H.~W.}\ \bibnamefont
  {Chan}},\ }\href@noop {} {\bibfield  {journal} {\bibinfo  {journal} {Phys.
  Rev. Lett.},\ }\textbf {\bibinfo {volume} {83}},\ \bibinfo {pages} {1187}
  (\bibinfo {year} {1999})}\BibitemShut {NoStop}%
\bibitem [{\citenamefont {Ganshin}\ \emph {et~al.}(2006)\citenamefont
  {Ganshin}, \citenamefont {Scheidemantel}, \citenamefont {Garcia},\ and\
  \citenamefont {Chan}}]{GanshinScheidemantelGarciaChan06}%
  \BibitemOpen
  \bibfield  {author} {\bibinfo {author} {\bibfnamefont {A.}~\bibnamefont
  {Ganshin}}, \bibinfo {author} {\bibfnamefont {S.}~\bibnamefont
  {Scheidemantel}}, \bibinfo {author} {\bibfnamefont {R.}~\bibnamefont
  {Garcia}}, \ and\ \bibinfo {author} {\bibfnamefont {M.~H.~W.}\ \bibnamefont
  {Chan}},\ }\href@noop {} {\bibfield  {journal} {\bibinfo  {journal} {Phys.
  Rev. Lett.},\ }\textbf {\bibinfo {volume} {97}},\ \bibinfo {pages} {075301}
  (\bibinfo {year} {2006})}\BibitemShut {NoStop}%
\bibitem [{\citenamefont {Fukuto}\ \emph {et~al.}(2005)\citenamefont {Fukuto},
  \citenamefont {Yano},\ and\ \citenamefont {Pershan}}]{FukutoYanoPershan05}%
  \BibitemOpen
  \bibfield  {author} {\bibinfo {author} {\bibfnamefont {M.}~\bibnamefont
  {Fukuto}}, \bibinfo {author} {\bibfnamefont {Y.~F.}\ \bibnamefont {Yano}}, \
  and\ \bibinfo {author} {\bibfnamefont {P.~S.}\ \bibnamefont {Pershan}},\
  }\href@noop {} {\bibfield  {journal} {\bibinfo  {journal} {Phys. Rev.
  Lett.},\ }\textbf {\bibinfo {volume} {94}},\ \bibinfo {pages} {135702}
  (\bibinfo {year} {2005})}\BibitemShut {NoStop}%
\bibitem [{\citenamefont {Hertlein}\ \emph {et~al.}(2008)\citenamefont
  {Hertlein}, \citenamefont {Helden}, \citenamefont {Gambassi}, \citenamefont
  {Dietrich},\ and\ \citenamefont
  {Bechinger}}]{HertleinHeldenGambassiDietrichBechinger08}%
  \BibitemOpen
  \bibfield  {author} {\bibinfo {author} {\bibfnamefont {C.}~\bibnamefont
  {Hertlein}}, \bibinfo {author} {\bibfnamefont {L.}~\bibnamefont {Helden}},
  \bibinfo {author} {\bibfnamefont {A.}~\bibnamefont {Gambassi}}, \bibinfo
  {author} {\bibfnamefont {S.}~\bibnamefont {Dietrich}}, \ and\ \bibinfo
  {author} {\bibfnamefont {C.}~\bibnamefont {Bechinger}},\ }\href@noop {}
  {\bibfield  {journal} {\bibinfo  {journal} {Nature},\ }\textbf {\bibinfo
  {volume} {451}},\ \bibinfo {pages} {172} (\bibinfo {year}
  {2008})}\BibitemShut {NoStop}%
\bibitem [{\citenamefont {Gambassi}\ \emph {et~al.}(2009)\citenamefont
  {Gambassi}, \citenamefont {Macio\l{}ek}, \citenamefont {Hertlein},
  \citenamefont {Nellen}, \citenamefont {Helden}, \citenamefont {Bechinger},\
  and\ \citenamefont {Dietrich}}]{Gambassi09}%
  \BibitemOpen
  \bibfield  {author} {\bibinfo {author} {\bibfnamefont {A.}~\bibnamefont
  {Gambassi}}, \bibinfo {author} {\bibfnamefont {A.}~\bibnamefont
  {Macio\l{}ek}}, \bibinfo {author} {\bibfnamefont {C.}~\bibnamefont
  {Hertlein}}, \bibinfo {author} {\bibfnamefont {U.}~\bibnamefont {Nellen}},
  \bibinfo {author} {\bibfnamefont {L.}~\bibnamefont {Helden}}, \bibinfo
  {author} {\bibfnamefont {C.}~\bibnamefont {Bechinger}}, \ and\ \bibinfo
  {author} {\bibfnamefont {S.}~\bibnamefont {Dietrich}},\ }\Doi
  {10.1103/PhysRevE.80.061143} {\bibfield  {journal} {\bibinfo  {journal}
  {Phys. Rev. E},\ }\textbf {\bibinfo {volume} {80}},\ \bibinfo {pages}
  {061143} (\bibinfo {year} {2009})}\BibitemShut {NoStop}%
\bibitem [{\citenamefont {Garcia}\ and\ \citenamefont
  {Chan}(2002)}]{GarciaChan02}%
  \BibitemOpen
  \bibfield  {author} {\bibinfo {author} {\bibfnamefont {R.}~\bibnamefont
  {Garcia}}\ and\ \bibinfo {author} {\bibfnamefont {M.~H.~W.}\ \bibnamefont
  {Chan}},\ }\href@noop {} {\bibfield  {journal} {\bibinfo  {journal} {Phys.
  Rev. Lett.},\ }\textbf {\bibinfo {volume} {88}},\ \bibinfo {pages} {086101}
  (\bibinfo {year} {2002})}\BibitemShut {NoStop}%
\bibitem [{\citenamefont {Krech}\ and\ \citenamefont
  {Dietrich}(1992)}]{KrechDietrich92b}%
  \BibitemOpen
  \bibfield  {author} {\bibinfo {author} {\bibfnamefont {M.}~\bibnamefont
  {Krech}}\ and\ \bibinfo {author} {\bibfnamefont {S.}~\bibnamefont
  {Dietrich}},\ }\href@noop {} {\bibfield  {journal} {\bibinfo  {journal}
  {Phys. Rev. A},\ }\textbf {\bibinfo {volume} {46}},\ \bibinfo {pages} {1886}
  (\bibinfo {year} {1992})}\BibitemShut {NoStop}%
\bibitem [{\citenamefont {Krech}(1994)}]{Krech94}%
  \BibitemOpen
  \bibfield  {author} {\bibinfo {author} {\bibfnamefont {M.}~\bibnamefont
  {Krech}},\ }\href@noop {} {\emph {\bibinfo {title} {Casimir Effect in
  Critical Systems}}}\ (\bibinfo  {publisher} {World Scientific},\ \bibinfo
  {address} {Singapore},\ \bibinfo {year} {1994})\BibitemShut {NoStop}%
\bibitem [{\citenamefont {Diehl}\ \emph {et~al.}(2006)\citenamefont {Diehl},
  \citenamefont {Grüneberg},\ and\ \citenamefont
  {Shpot}}]{DiehlGruenebergShpot06}%
  \BibitemOpen
  \bibfield  {author} {\bibinfo {author} {\bibfnamefont {H.~W.}\ \bibnamefont
  {Diehl}}, \bibinfo {author} {\bibfnamefont {D.}~\bibnamefont {Grüneberg}}, \
  and\ \bibinfo {author} {\bibfnamefont {M.~A.}\ \bibnamefont {Shpot}},\
  }\href@noop {} {\bibfield  {journal} {\bibinfo  {journal} {Europhys. Lett.},\
  }\textbf {\bibinfo {volume} {75}},\ \bibinfo {pages} {241} (\bibinfo {year}
  {2006})}\BibitemShut {NoStop}%
\bibitem [{\citenamefont {Gr\"{u}neberg}\ and\ \citenamefont
  {Diehl}(2008)}]{GruenebergDiehl08}%
  \BibitemOpen
  \bibfield  {author} {\bibinfo {author} {\bibfnamefont {D.}~\bibnamefont
  {Gr\"{u}neberg}}\ and\ \bibinfo {author} {\bibfnamefont {H.~W.}\ \bibnamefont
  {Diehl}},\ }\Doi {10.1103/PhysRevB.77.115409} {\bibfield  {journal} {\bibinfo
   {journal} {Phys. Rev.~B},\ }\textbf {\bibinfo {volume} {77}},\ \bibinfo
  {eid} {115409} (\bibinfo {year} {2008})}\BibitemShut {NoStop}%
\bibitem [{\citenamefont {Macio{\l}ek}\ \emph {et~al.}(2007)\citenamefont
  {Macio{\l}ek}, \citenamefont {Gambassi},\ and\ \citenamefont
  {Dietrich}}]{MaciolekGambassiDietrich07}%
  \BibitemOpen
  \bibfield  {author} {\bibinfo {author} {\bibfnamefont {A.}~\bibnamefont
  {Macio{\l}ek}}, \bibinfo {author} {\bibfnamefont {A.}~\bibnamefont
  {Gambassi}}, \ and\ \bibinfo {author} {\bibfnamefont {S.}~\bibnamefont
  {Dietrich}},\ }\href@noop {} {\bibfield  {journal} {\bibinfo  {journal}
  {Phys. Rev.~E},\ }\textbf {\bibinfo {volume} {76}},\ \bibinfo {pages}
  {031124} (\bibinfo {year} {2007})}\BibitemShut {NoStop}%
\bibitem [{\citenamefont {Zandi}\ \emph {et~al.}(2007)\citenamefont {Zandi},
  \citenamefont {Shackell}, \citenamefont {Rudnick}, \citenamefont {Kardar},\
  and\ \citenamefont {Chayes}}]{ZandiShackellRudnickKardarChayes07}%
  \BibitemOpen
  \bibfield  {author} {\bibinfo {author} {\bibfnamefont {R.}~\bibnamefont
  {Zandi}}, \bibinfo {author} {\bibfnamefont {A.}~\bibnamefont {Shackell}},
  \bibinfo {author} {\bibfnamefont {J.}~\bibnamefont {Rudnick}}, \bibinfo
  {author} {\bibfnamefont {M.}~\bibnamefont {Kardar}}, \ and\ \bibinfo {author}
  {\bibfnamefont {L.~P.}\ \bibnamefont {Chayes}},\ }\href@noop {} {\bibfield
  {journal} {\bibinfo  {journal} {Phys. Rev.~E},\ }\textbf {\bibinfo {volume}
  {76}},\ \bibinfo {pages} {030601} (\bibinfo {year} {2007})}\BibitemShut
  {NoStop}%
\bibitem [{\citenamefont {Li}\ and\ \citenamefont {Kardar}(1991)}]{LiKardar91}%
  \BibitemOpen
  \bibfield  {author} {\bibinfo {author} {\bibfnamefont {H.}~\bibnamefont
  {Li}}\ and\ \bibinfo {author} {\bibfnamefont {M.}~\bibnamefont {Kardar}},\
  }\href@noop {} {\bibfield  {journal} {\bibinfo  {journal} {Phys. Rev.
  Lett.},\ }\textbf {\bibinfo {volume} {67}},\ \bibinfo {pages} {3275}
  (\bibinfo {year} {1991})}\BibitemShut {NoStop}%
\bibitem [{\citenamefont {Li}\ and\ \citenamefont {Kardar}(1992)}]{LiKardar92}%
  \BibitemOpen
  \bibfield  {author} {\bibinfo {author} {\bibfnamefont {H.}~\bibnamefont
  {Li}}\ and\ \bibinfo {author} {\bibfnamefont {M.}~\bibnamefont {Kardar}},\
  }\href@noop {} {\bibfield  {journal} {\bibinfo  {journal} {Phys. Rev.~A},\
  }\textbf {\bibinfo {volume} {46}},\ \bibinfo {pages} {6490} (\bibinfo {year}
  {1992})}\BibitemShut {NoStop}%
\bibitem [{\citenamefont {Kardar}\ and\ \citenamefont
  {Golestanian}(1999)}]{KardarGolestanian99}%
  \BibitemOpen
  \bibfield  {author} {\bibinfo {author} {\bibfnamefont {M.}~\bibnamefont
  {Kardar}}\ and\ \bibinfo {author} {\bibfnamefont {R.}~\bibnamefont
  {Golestanian}},\ }\href@noop {} {\bibfield  {journal} {\bibinfo  {journal}
  {Rev. Mod. Phys.},\ }\textbf {\bibinfo {volume} {71}},\ \bibinfo {pages}
  {1233} (\bibinfo {year} {1999})}\BibitemShut {NoStop}%
\bibitem [{\citenamefont {Zandi}\ \emph {et~al.}(2004)\citenamefont {Zandi},
  \citenamefont {Rudnick},\ and\ \citenamefont
  {Kardar}}]{ZandiRudnickKardar04}%
  \BibitemOpen
  \bibfield  {author} {\bibinfo {author} {\bibfnamefont {R.}~\bibnamefont
  {Zandi}}, \bibinfo {author} {\bibfnamefont {J.}~\bibnamefont {Rudnick}}, \
  and\ \bibinfo {author} {\bibfnamefont {M.}~\bibnamefont {Kardar}},\
  }\href@noop {} {\bibfield  {journal} {\bibinfo  {journal} {Phys. Rev.
  Lett.},\ }\textbf {\bibinfo {volume} {93}},\ \bibinfo {pages} {155302}
  (\bibinfo {year} {2004})}\BibitemShut {NoStop}%
\bibitem [{\citenamefont {Hucht}(2007)}]{Hucht07a}%
  \BibitemOpen
  \bibfield  {author} {\bibinfo {author} {\bibfnamefont {A.}~\bibnamefont
  {Hucht}},\ }\Doi {10.1103/PhysRevLett.99.185301} {\bibfield  {journal}
  {\bibinfo  {journal} {Phys. Rev. Lett.},\ }\textbf {\bibinfo {volume} {99}},\
  \bibinfo {pages} {185301} (\bibinfo {year} {2007})}\BibitemShut {NoStop}%
\bibitem [{\citenamefont {Dantchev}\ and\ \citenamefont
  {Krech}(2004)}]{DantchevKrech04}%
  \BibitemOpen
  \bibfield  {author} {\bibinfo {author} {\bibfnamefont {D.}~\bibnamefont
  {Dantchev}}\ and\ \bibinfo {author} {\bibfnamefont {M.}~\bibnamefont
  {Krech}},\ }\href@noop {} {\bibfield  {journal} {\bibinfo  {journal} {Phys.
  Rev.~E},\ }\textbf {\bibinfo {volume} {69}},\ \bibinfo {pages} {046119}
  (\bibinfo {year} {2004})}\BibitemShut {NoStop}%
\bibitem [{\citenamefont {Hasenbusch}(2010){\natexlab{a}}}]{Hasenbusch0907}%
  \BibitemOpen
  \bibfield  {author} {\bibinfo {author} {\bibfnamefont {M.}~\bibnamefont
  {Hasenbusch}},\ }\Doi {10.1103/PhysRevB.81.165412} {\bibfield  {journal}
  {\bibinfo  {journal} {Phys. Rev.~B},\ }\textbf {\bibinfo {volume} {81}},\
  \bibinfo {pages} {165412} (\bibinfo {year} {2010}{\natexlab{a}})},\ \bibinfo
  {note} {arXiv:0907.2847}\BibitemShut {NoStop}%
\bibitem [{\citenamefont {Hasenbusch}(2010){\natexlab{b}}}]{Hasenbusch1005}%
  \BibitemOpen
  \bibfield  {author} {\bibinfo {author} {\bibfnamefont {M.}~\bibnamefont
  {Hasenbusch}},\ }\Doi {10.1103/PhysRevB.82.104425} {\bibfield  {journal}
  {\bibinfo  {journal} {Phys. Rev.~B},\ }\textbf {\bibinfo {volume} {82}},\
  \bibinfo {pages} {104425} (\bibinfo {year} {2010}{\natexlab{b}})},\ \bibinfo
  {note} {arXiv:1005.4749}\BibitemShut {NoStop}%
\bibitem [{\citenamefont {Vasilyev}\ \emph {et~al.}(2007)\citenamefont
  {Vasilyev}, \citenamefont {Gambassi}, \citenamefont {Macio{\l}ek},\ and\
  \citenamefont {Dietrich}}]{VasilyevGambassiMaciolekDietrich07}%
  \BibitemOpen
  \bibfield  {author} {\bibinfo {author} {\bibfnamefont {O.}~\bibnamefont
  {Vasilyev}}, \bibinfo {author} {\bibfnamefont {A.}~\bibnamefont {Gambassi}},
  \bibinfo {author} {\bibfnamefont {A.}~\bibnamefont {Macio{\l}ek}}, \ and\
  \bibinfo {author} {\bibfnamefont {S.}~\bibnamefont {Dietrich}},\ }\href@noop
  {} {\bibfield  {journal} {\bibinfo  {journal} {Europhys. Lett.},\ }\textbf
  {\bibinfo {volume} {80}},\ \bibinfo {pages} {60009} (\bibinfo {year}
  {2007})}\BibitemShut {NoStop}%
\bibitem [{\citenamefont {Hasenbusch}(2009){\natexlab{a}}}]{Hasenbusch0905}%
  \BibitemOpen
  \bibfield  {author} {\bibinfo {author} {\bibfnamefont {M.}~\bibnamefont
  {Hasenbusch}},\ }\Doi {10.1088/1742-5468/2009/07/P07031} {\bibfield
  {journal} {\bibinfo  {journal} {J.~Stat. Mech.: Theory Exp.},\ \bibinfo
  {pages} {P07031}} (\bibinfo {year} {2009}{\natexlab{a}})},\ \bibinfo {note}
  {arXiv:0905.2096}\BibitemShut {NoStop}%
\bibitem [{\citenamefont {Hasenbusch}(2009){\natexlab{b}}}]{Hasenbusch0908}%
  \BibitemOpen
  \bibfield  {author} {\bibinfo {author} {\bibfnamefont {M.}~\bibnamefont
  {Hasenbusch}},\ }\Doi {10.1103/PhysRevE.80.061120} {\bibfield  {journal}
  {\bibinfo  {journal} {Phys. Rev.~E},\ }\textbf {\bibinfo {volume} {80}},\
  \bibinfo {pages} {061120} (\bibinfo {year} {2009}{\natexlab{b}})},\ \bibinfo
  {note} {arXiv:0908.3582}\BibitemShut {NoStop}%
\bibitem [{\citenamefont {Vasilyev}\ \emph {et~al.}(2009)\citenamefont
  {Vasilyev}, \citenamefont {Gambassi}, \citenamefont {Macio{\l}ek},\ and\
  \citenamefont {Dietrich}}]{VasilyevGambassiMaciolekDietrich09}%
  \BibitemOpen
  \bibfield  {author} {\bibinfo {author} {\bibfnamefont {O.}~\bibnamefont
  {Vasilyev}}, \bibinfo {author} {\bibfnamefont {A.}~\bibnamefont {Gambassi}},
  \bibinfo {author} {\bibfnamefont {A.}~\bibnamefont {Macio{\l}ek}}, \ and\
  \bibinfo {author} {\bibfnamefont {S.}~\bibnamefont {Dietrich}},\ }\Doi
  {10.1103/PhysRevE.79.041142} {\bibfield  {journal} {\bibinfo  {journal}
  {Phys. Rev.~E},\ }\textbf {\bibinfo {volume} {79}},\ \bibinfo {eid} {041142}
  (\bibinfo {year} {2009})}\BibitemShut {NoStop}%
\bibitem [{\citenamefont {Toldin}\ and\ \citenamefont
  {Dietrich}(2010)}]{ToldinDietrich10}%
  \BibitemOpen
  \bibfield  {author} {\bibinfo {author} {\bibfnamefont {F.~P.}\ \bibnamefont
  {Toldin}}\ and\ \bibinfo {author} {\bibfnamefont {S.}~\bibnamefont
  {Dietrich}},\ }\href {http://stacks.iop.org/1742-5468/2010/i=11/a=P11003}
  {\bibfield  {journal} {\bibinfo  {journal} {J.~Stat. Mech.: Theory Exp.},\
  }\textbf {\bibinfo {volume} {2010}},\ \bibinfo {pages} {P11003} (\bibinfo
  {year} {2010})}\BibitemShut {NoStop}%
\bibitem [{\citenamefont {Dohm}(2009)}]{Dohm09}%
  \BibitemOpen
  \bibfield  {author} {\bibinfo {author} {\bibfnamefont {V.}~\bibnamefont
  {Dohm}},\ }\href@noop {} {\bibfield  {journal} {\bibinfo  {journal}
  {Europhys. Lett.},\ }\textbf {\bibinfo {volume} {86}},\ \bibinfo {pages}
  {20001 (5pp)} (\bibinfo {year} {2009})}\BibitemShut {NoStop}%
\bibitem [{\citenamefont {Privman}(1990)}]{Privman90}%
  \BibitemOpen
  \bibfield  {author} {\bibinfo {author} {\bibfnamefont {V.}~\bibnamefont
  {Privman}},\ }in\ \href@noop {} {\emph {\bibinfo {booktitle} {Finite Size
  Scaling and Numerical Simulation of Statistical Systems}}},\ \bibinfo
  {editor} {edited by\ \bibinfo {editor} {\bibfnamefont {V.}~\bibnamefont
  {Privman}}}\ (\bibinfo  {publisher} {World Scientific},\ \bibinfo {address}
  {Singapore},\ \bibinfo {year} {1990})\ Chap.~\bibinfo {chapter}
  {1}\BibitemShut {NoStop}%
\bibitem [{Note1()}]{Note1}%
  \BibitemOpen
  \bibinfo {note} {Throughout this work, the symbol $\sim $ means
  {}``asymptotically equal'' in the respective limit, $L_{\parallel },L_{\perp
  }\rightarrow \infty $, $T\rightarrow T_{\protect \mathrm {c}}$, keeping the
  scaling variables $x$ and $\rho $ fixed, i.\protect \tmspace +\thinmuskip
  {.1667em}e., $f(L)\sim g(L)\Leftrightarrow \protect \qopname \relax
  m{lim}_{L\rightarrow \infty }f(L)/g(L)=1.$}\BibitemShut {NoStop}%
\bibitem [{\citenamefont {Campostrini}\ \emph {et~al.}(1999)\citenamefont
  {Campostrini}, \citenamefont {Pelissetto}, \citenamefont {Rossi},\ and\
  \citenamefont {Vicari}}]{CampostriniPelissettoRossiVicari99}%
  \BibitemOpen
  \bibfield  {author} {\bibinfo {author} {\bibfnamefont {M.}~\bibnamefont
  {Campostrini}}, \bibinfo {author} {\bibfnamefont {A.}~\bibnamefont
  {Pelissetto}}, \bibinfo {author} {\bibfnamefont {P.}~\bibnamefont {Rossi}}, \
  and\ \bibinfo {author} {\bibfnamefont {E.}~\bibnamefont {Vicari}},\ }\Doi
  {10.1103/PhysRevE.60.3526} {\bibfield  {journal} {\bibinfo  {journal} {Phys.
  Rev.~E},\ }\textbf {\bibinfo {volume} {60}},\ \bibinfo {pages} {3526}
  (\bibinfo {year} {1999})}\BibitemShut {NoStop}%
\bibitem [{\citenamefont {Butera}\ and\ \citenamefont
  {Comi}(2002)}]{ButeraComi02}%
  \BibitemOpen
  \bibfield  {author} {\bibinfo {author} {\bibfnamefont {P.}~\bibnamefont
  {Butera}}\ and\ \bibinfo {author} {\bibfnamefont {M.}~\bibnamefont {Comi}},\
  }\Doi {10.1103/PhysRevB.65.144431} {\bibfield  {journal} {\bibinfo  {journal}
  {Phys. Rev.~B},\ }\textbf {\bibinfo {volume} {65}},\ \bibinfo {pages}
  {144431} (\bibinfo {year} {2002})}\BibitemShut {NoStop}%
\bibitem [{\citenamefont {Fisher}(1971)}]{Fisher71}%
  \BibitemOpen
  \bibfield  {author} {\bibinfo {author} {\bibfnamefont {M.~E.}\ \bibnamefont
  {Fisher}},\ }in\ \href@noop {} {\emph {\bibinfo {booktitle} {Critical
  Phenomena, Proceedings of the 51st Enrico Fermi Summer School, Varenna,
  Italy}}},\ \bibinfo {editor} {edited by\ \bibinfo {editor} {\bibfnamefont
  {M.~S.}\ \bibnamefont {Green}}}\ (\bibinfo  {publisher} {Academic Press},\
  \bibinfo {address} {New York},\ \bibinfo {year} {1971})\ pp.\ \bibinfo
  {pages} {73--98}\BibitemShut {NoStop}%
\bibitem [{\citenamefont {Dantchev}\ \emph {et~al.}(2006)\citenamefont
  {Dantchev}, \citenamefont {Diehl},\ and\ \citenamefont
  {Gr\"{u}neberg}}]{DantchevDiehlGrueneberg06}%
  \BibitemOpen
  \bibfield  {author} {\bibinfo {author} {\bibfnamefont {D.}~\bibnamefont
  {Dantchev}}, \bibinfo {author} {\bibfnamefont {H.~W.}\ \bibnamefont {Diehl}},
  \ and\ \bibinfo {author} {\bibfnamefont {D.}~\bibnamefont {Gr\"{u}neberg}},\
  }\Doi {10.1103/PhysRevE.73.016131} {\bibfield  {journal} {\bibinfo  {journal}
  {Phys. Rev.~E},\ }\textbf {\bibinfo {volume} {73}},\ \bibinfo {eid} {016131}
  (\bibinfo {year} {2006})}\BibitemShut {NoStop}%
\bibitem [{\citenamefont {Gr{\"u}neberg}\ and\ \citenamefont
  {Hucht}(2004)}]{GruenebergHucht04}%
  \BibitemOpen
  \bibfield  {author} {\bibinfo {author} {\bibfnamefont {D.}~\bibnamefont
  {Gr{\"u}neberg}}\ and\ \bibinfo {author} {\bibfnamefont {A.}~\bibnamefont
  {Hucht}},\ }\Doi {10.1103/PhysRevE.69.036104} {\bibfield  {journal} {\bibinfo
   {journal} {Phys. Rev.~E},\ }\textbf {\bibinfo {volume} {69}},\ \bibinfo
  {pages} {036104} (\bibinfo {year} {2004})}\BibitemShut {NoStop}%
\bibitem [{\citenamefont {Danchev}(1998)}]{Dantchev98}%
  \BibitemOpen
  \bibfield  {author} {\bibinfo {author} {\bibfnamefont {D.~M.}\ \bibnamefont
  {Danchev}},\ }\Doi {10.1103/PhysRevE.58.1455} {\bibfield  {journal} {\bibinfo
   {journal} {Phys. Rev.~E},\ }\textbf {\bibinfo {volume} {58}},\ \bibinfo
  {pages} {1455} (\bibinfo {year} {1998})}\BibitemShut {NoStop}%
\bibitem [{\citenamefont {Brankov}\ \emph {et~al.}(2000)\citenamefont
  {Brankov}, \citenamefont {Dantchev},\ and\ \citenamefont
  {Tonchev}}]{BrankovDantchevTonchev00}%
  \BibitemOpen
  \bibfield  {author} {\bibinfo {author} {\bibfnamefont {J.~G.}\ \bibnamefont
  {Brankov}}, \bibinfo {author} {\bibfnamefont {D.~M.}\ \bibnamefont
  {Dantchev}}, \ and\ \bibinfo {author} {\bibfnamefont {N.~S.}\ \bibnamefont
  {Tonchev}},\ }\href@noop {} {\emph {\bibinfo {title} {Theory of Critical
  Phenomena in Finite-Size Systems -- Scaling and Quantum Effects}}}\ (\bibinfo
   {publisher} {World Scientific},\ \bibinfo {address} {Singapore},\ \bibinfo
  {year} {2000})\BibitemShut {NoStop}%
\bibitem [{\citenamefont {Dantchev}\ \emph {et~al.}(2003)\citenamefont
  {Dantchev}, \citenamefont {Krech},\ and\ \citenamefont
  {Dietrich}}]{DantchevKrechDietrich03}%
  \BibitemOpen
  \bibfield  {author} {\bibinfo {author} {\bibfnamefont {D.}~\bibnamefont
  {Dantchev}}, \bibinfo {author} {\bibfnamefont {M.}~\bibnamefont {Krech}}, \
  and\ \bibinfo {author} {\bibfnamefont {S.}~\bibnamefont {Dietrich}},\
  }\href@noop {} {\bibfield  {journal} {\bibinfo  {journal} {Phys. Rev.~E},\
  }\textbf {\bibinfo {volume} {67}},\ \bibinfo {pages} {066120} (\bibinfo
  {year} {2003})}\BibitemShut {NoStop}%
\bibitem [{\citenamefont {Dantchev}\ and\ \citenamefont
  {Gr\"uneberg}(2009)}]{DantchevGrueneberg09}%
  \BibitemOpen
  \bibfield  {author} {\bibinfo {author} {\bibfnamefont {D.}~\bibnamefont
  {Dantchev}}\ and\ \bibinfo {author} {\bibfnamefont {D.}~\bibnamefont
  {Gr\"uneberg}},\ }\Doi {10.1103/PhysRevE.79.041103} {\bibfield  {journal}
  {\bibinfo  {journal} {Phys. Rev.~E},\ }\textbf {\bibinfo {volume} {79}},\
  \bibinfo {pages} {041103} (\bibinfo {year} {2009})}\BibitemShut {NoStop}%
\bibitem [{\citenamefont {Krech}\ and\ \citenamefont
  {Landau}(1996)}]{KrechLandau96}%
  \BibitemOpen
  \bibfield  {author} {\bibinfo {author} {\bibfnamefont {M.}~\bibnamefont
  {Krech}}\ and\ \bibinfo {author} {\bibfnamefont {D.~P.}\ \bibnamefont
  {Landau}},\ }\Doi {10.1103/PhysRevE.53.4414} {\bibfield  {journal} {\bibinfo
  {journal} {Phys. Rev.~E},\ }\textbf {\bibinfo {volume} {53}},\ \bibinfo
  {pages} {4414} (\bibinfo {year} {1996})}\BibitemShut {NoStop}%
\bibitem [{\citenamefont {Schmidt}\ and\ \citenamefont
  {Diehl}(2008)}]{SchmidtDiehl08}%
  \BibitemOpen
  \bibfield  {author} {\bibinfo {author} {\bibfnamefont {F.~M.}\ \bibnamefont
  {Schmidt}}\ and\ \bibinfo {author} {\bibfnamefont {H.~W.}\ \bibnamefont
  {Diehl}},\ }\Doi {10.1103/PhysRevLett.101.100601} {\bibfield  {journal}
  {\bibinfo  {journal} {Phys. Rev. Lett.},\ }\textbf {\bibinfo {volume}
  {101}},\ \bibinfo {pages} {100601} (\bibinfo {year} {2008})}\BibitemShut
  {NoStop}%
\bibitem [{\citenamefont {Diehl}\ and\ \citenamefont
  {Gr\"uneberg}(2009)}]{DiehlGrueneberg09}%
  \BibitemOpen
  \bibfield  {author} {\bibinfo {author} {\bibfnamefont {H.~W.}\ \bibnamefont
  {Diehl}}\ and\ \bibinfo {author} {\bibfnamefont {D.}~\bibnamefont
  {Gr\"uneberg}},\ }\Doi {10.1016/j.nuclphysb.2009.07.010} {\bibfield
  {journal} {\bibinfo  {journal} {Nucl. Phys.~B},\ }\textbf {\bibinfo {volume}
  {822}},\ \bibinfo {pages} {517 } (\bibinfo {year} {2009})}\BibitemShut
  {NoStop}%
\bibitem [{\citenamefont {Burgsm\"uller}\ \emph {et~al.}(2010)\citenamefont
  {Burgsm\"uller}, \citenamefont {Diehl},\ and\ \citenamefont
  {Shpot}}]{BurgsmuellerDiehlShpot10}%
  \BibitemOpen
  \bibfield  {author} {\bibinfo {author} {\bibfnamefont {M.}~\bibnamefont
  {Burgsm\"uller}}, \bibinfo {author} {\bibfnamefont {H.~W.}\ \bibnamefont
  {Diehl}}, \ and\ \bibinfo {author} {\bibfnamefont {M.~A.}\ \bibnamefont
  {Shpot}},\ }\href {http://stacks.iop.org/1742-5468/2010/i=11/a=P11020}
  {\bibfield  {journal} {\bibinfo  {journal} {J.~Stat. Mech.: Theory Exp.},\
  \bibinfo {pages} {P11020}} (\bibinfo {year} {2010})}\BibitemShut {NoStop}%
\bibitem [{\citenamefont {Bhanot}\ \emph {et~al.}(1994)\citenamefont {Bhanot},
  \citenamefont {Creutz}, \citenamefont {Horvath}, \citenamefont {Lacki},\ and\
  \citenamefont {Weckel}}]{BhanotCreutzHorvathLackiWeckel94}%
  \BibitemOpen
  \bibfield  {author} {\bibinfo {author} {\bibfnamefont {G.}~\bibnamefont
  {Bhanot}}, \bibinfo {author} {\bibfnamefont {M.}~\bibnamefont {Creutz}},
  \bibinfo {author} {\bibfnamefont {I.}~\bibnamefont {Horvath}}, \bibinfo
  {author} {\bibfnamefont {J.}~\bibnamefont {Lacki}}, \ and\ \bibinfo {author}
  {\bibfnamefont {J.}~\bibnamefont {Weckel}},\ }\Doi {10.1103/PhysRevE.49.2445}
  {\bibfield  {journal} {\bibinfo  {journal} {Phys. Rev.~E},\ }\textbf
  {\bibinfo {volume} {49}},\ \bibinfo {pages} {2445} (\bibinfo {year}
  {1994})}\BibitemShut {NoStop}%
\bibitem [{\citenamefont {Feng}\ and\ \citenamefont
  {Bl\"ote}(2010)}]{FengBloete10}%
  \BibitemOpen
  \bibfield  {author} {\bibinfo {author} {\bibfnamefont {X.}~\bibnamefont
  {Feng}}\ and\ \bibinfo {author} {\bibfnamefont {H.~W.~J.}\ \bibnamefont
  {Bl\"ote}},\ }\Doi {10.1103/PhysRevE.81.031103} {\bibfield  {journal}
  {\bibinfo  {journal} {Phys. Rev.~E},\ }\textbf {\bibinfo {volume} {81}},\
  \bibinfo {pages} {031103} (\bibinfo {year} {2010})}\BibitemShut {NoStop}%
\bibitem [{\citenamefont {Arisue}\ and\ \citenamefont
  {Fujiwara}(2003)}]{ArisueFujiwara03a}%
  \BibitemOpen
  \bibfield  {author} {\bibinfo {author} {\bibfnamefont {H.}~\bibnamefont
  {Arisue}}\ and\ \bibinfo {author} {\bibfnamefont {T.}~\bibnamefont
  {Fujiwara}},\ }\Doi {10.1103/PhysRevE.67.066109} {\bibfield  {journal}
  {\bibinfo  {journal} {Phys. Rev.~E},\ }\textbf {\bibinfo {volume} {67}},\
  \bibinfo {pages} {066109} (\bibinfo {year} {2003})},\ \bibinfo {note} {there
  is a typo in the 42th order term, the correct value appears in
  arXiv:hep-lat/0209002}\BibitemShut {NoStop}%
\bibitem [{\citenamefont {Wolff}(1989)}]{Wolff89}%
  \BibitemOpen
  \bibfield  {author} {\bibinfo {author} {\bibfnamefont {U.}~\bibnamefont
  {Wolff}},\ }\href@noop {} {\bibfield  {journal} {\bibinfo  {journal} {Phys.
  Rev. Lett.},\ }\textbf {\bibinfo {volume} {62}},\ \bibinfo {pages} {361}
  (\bibinfo {year} {1989})}\BibitemShut {NoStop}%
\bibitem [{\citenamefont {Deng}\ and\ \citenamefont
  {Bl\"ote}(2003)}]{DengBloete03}%
  \BibitemOpen
  \bibfield  {author} {\bibinfo {author} {\bibfnamefont {Y.}~\bibnamefont
  {Deng}}\ and\ \bibinfo {author} {\bibfnamefont {H.~W.~J.}\ \bibnamefont
  {Bl\"ote}},\ }\Doi {10.1103/PhysRevE.68.036125} {\bibfield  {journal}
  {\bibinfo  {journal} {Phys. Rev.~E},\ }\textbf {\bibinfo {volume} {68}},\
  \bibinfo {pages} {036125} (\bibinfo {year} {2003})}\BibitemShut {NoStop}%
\bibitem [{\citenamefont {Kitatani}\ \emph {et~al.}(1996)\citenamefont
  {Kitatani}, \citenamefont {Ohta},\ and\ \citenamefont
  {Ito}}]{KitataniOhtaIto96}%
  \BibitemOpen
  \bibfield  {author} {\bibinfo {author} {\bibfnamefont {H.}~\bibnamefont
  {Kitatani}}, \bibinfo {author} {\bibfnamefont {M.}~\bibnamefont {Ohta}}, \
  and\ \bibinfo {author} {\bibfnamefont {N.}~\bibnamefont {Ito}},\ }\Doi
  {10.1143/JPSJ.65.4050} {\bibfield  {journal} {\bibinfo  {journal} {J.~Phys.
  Soc. Jpn.},\ }\textbf {\bibinfo {volume} {65}},\ \bibinfo {pages} {4050}
  (\bibinfo {year} {1996})}\BibitemShut {NoStop}%
\bibitem [{\citenamefont {Caselle}\ and\ \citenamefont
  {Hasenbusch}(1996)}]{CaselleHasenbusch96}%
  \BibitemOpen
  \bibfield  {author} {\bibinfo {author} {\bibfnamefont {M.}~\bibnamefont
  {Caselle}}\ and\ \bibinfo {author} {\bibfnamefont {M.}~\bibnamefont
  {Hasenbusch}},\ }\Doi {DOI: 10.1016/0550-3213(96)00161-7} {\bibfield
  {journal} {\bibinfo  {journal} {Nucl. Phys.~B},\ }\textbf {\bibinfo {volume}
  {470}},\ \bibinfo {pages} {435 } (\bibinfo {year} {1996})},\ ISSN \bibinfo
  {issn} {0550-3213}\BibitemShut {NoStop}%
\bibitem [{\citenamefont {Krech}(1997)}]{Krech97}%
  \BibitemOpen
  \bibfield  {author} {\bibinfo {author} {\bibfnamefont {M.}~\bibnamefont
  {Krech}},\ }\Doi {10.1103/PhysRevE.56.1642} {\bibfield  {journal} {\bibinfo
  {journal} {Phys. Rev.~E},\ }\textbf {\bibinfo {volume} {56}},\ \bibinfo
  {pages} {1642} (\bibinfo {year} {1997})}\BibitemShut {NoStop}%
\bibitem [{Note2()}]{Note2}%
  \BibitemOpen
  \bibinfo {note} {At $\rho =1$ all quantities $q$ obey $q=q_{\perp
  }=q_{\parallel }$}\BibitemShut {NoStop}%
\bibitem [{\citenamefont {Bachas}(2007)}]{Bachas07}%
  \BibitemOpen
  \bibfield  {author} {\bibinfo {author} {\bibfnamefont {C.~P.}\ \bibnamefont
  {Bachas}},\ }\href {http://stacks.iop.org/1751-8121/40/i=30/a=028} {\bibfield
   {journal} {\bibinfo  {journal} {J.~Phys A: Math. Theor.},\ }\textbf
  {\bibinfo {volume} {40}},\ \bibinfo {pages} {9089} (\bibinfo {year}
  {2007})}\BibitemShut {NoStop}%
\bibitem [{\citenamefont {Privman}\ and\ \citenamefont
  {Fisher}(1983)}]{PrivmanFisher83}%
  \BibitemOpen
  \bibfield  {author} {\bibinfo {author} {\bibfnamefont {V.}~\bibnamefont
  {Privman}}\ and\ \bibinfo {author} {\bibfnamefont {M.~E.}\ \bibnamefont
  {Fisher}},\ }\href@noop {} {\bibfield  {journal} {\bibinfo  {journal}
  {J.~Stat. Phys.},\ }\textbf {\bibinfo {volume} {33}},\ \bibinfo {pages} {385}
  (\bibinfo {year} {1983})}\BibitemShut {NoStop}%
\bibitem [{\citenamefont {Mon}(1985)}]{Mon85}%
  \BibitemOpen
  \bibfield  {author} {\bibinfo {author} {\bibfnamefont {K.~K.}\ \bibnamefont
  {Mon}},\ }\Doi {10.1103/PhysRevLett.54.2671} {\bibfield  {journal} {\bibinfo
  {journal} {Phys. Rev. Lett.},\ }\textbf {\bibinfo {volume} {54}},\ \bibinfo
  {pages} {2671} (\bibinfo {year} {1985})}\BibitemShut {NoStop}%
\bibitem [{\citenamefont {Ferdinand}\ and\ \citenamefont
  {Fisher}(1969)}]{FerdinandFisher69}%
  \BibitemOpen
  \bibfield  {author} {\bibinfo {author} {\bibfnamefont {A.~E.}\ \bibnamefont
  {Ferdinand}}\ and\ \bibinfo {author} {\bibfnamefont {M.~E.}\ \bibnamefont
  {Fisher}},\ }\href@noop {} {\bibfield  {journal} {\bibinfo  {journal} {Phys.
  Rev.},\ }\textbf {\bibinfo {volume} {185}},\ \bibinfo {pages} {832} (\bibinfo
  {year} {1969})},\ \bibinfo {note} {there is a typo in Eq. (3.36), the term
  $\xi S_{1}(n)\tau^{2}/2$ is missing}\BibitemShut {NoStop}%
\bibitem [{\citenamefont {Kaufman}(1949)}]{Kaufman49}%
  \BibitemOpen
  \bibfield  {author} {\bibinfo {author} {\bibfnamefont {B.}~\bibnamefont
  {Kaufman}},\ }\Doi {10.1103/PhysRev.76.1232} {\bibfield  {journal} {\bibinfo
  {journal} {Phys. Rev.},\ }\textbf {\bibinfo {volume} {76}},\ \bibinfo {pages}
  {1232} (\bibinfo {year} {1949})}\BibitemShut {NoStop}%
\bibitem [{\citenamefont {{Wolfram Research, Inc.}}(2008)}]{MMA7}%
  \BibitemOpen
  \bibfield  {author} {\bibinfo {author} {\bibnamefont {{Wolfram Research,
  Inc.}}},\ }\href@noop {} {\emph {\bibinfo {title} {Mathematica V7.0}}},\
  \bibinfo {address} {Champaign, Illinois} (\bibinfo {year} {2008})\BibitemShut
  {NoStop}%
\bibitem [{\citenamefont {Whittaker}\ and\ \citenamefont
  {Watson}(1990)}]{WhittakerWatson90}%
  \BibitemOpen
  \bibfield  {author} {\bibinfo {author} {\bibfnamefont {E.~T.}\ \bibnamefont
  {Whittaker}}\ and\ \bibinfo {author} {\bibfnamefont {G.~N.}\ \bibnamefont
  {Watson}},\ }\href@noop {} {\emph {\bibinfo {title} {A Course in Modern
  Analysis}}},\ \bibinfo {edition} {4th}\ ed.\ (\bibinfo  {publisher}
  {Cambridge University Press},\ \bibinfo {year} {1990})\BibitemShut {NoStop}%
\bibitem [{\citenamefont {Rudnick}\ \emph {et~al.}(2010)\citenamefont
  {Rudnick}, \citenamefont {Zandi}, \citenamefont {Shackell},\ and\
  \citenamefont {Abraham}}]{RudnickZandiShackellAbraham10}%
  \BibitemOpen
  \bibfield  {author} {\bibinfo {author} {\bibfnamefont {J.}~\bibnamefont
  {Rudnick}}, \bibinfo {author} {\bibfnamefont {R.}~\bibnamefont {Zandi}},
  \bibinfo {author} {\bibfnamefont {A.}~\bibnamefont {Shackell}}, \ and\
  \bibinfo {author} {\bibfnamefont {D.}~\bibnamefont {Abraham}},\ }\Doi
  {10.1103/PhysRevE.82.041118} {\bibfield  {journal} {\bibinfo  {journal}
  {Phys. Rev.~E},\ }\textbf {\bibinfo {volume} {82}},\ \bibinfo {pages}
  {041118} (\bibinfo {year} {2010})}\BibitemShut {NoStop}%
\bibitem [{\citenamefont {Hucht}(2002)}]{Hucht02a}%
  \BibitemOpen
  \bibfield  {author} {\bibinfo {author} {\bibfnamefont {A.}~\bibnamefont
  {Hucht}},\ }\href@noop {} {\bibfield  {journal} {\bibinfo  {journal} {J.~Phys
  A: Math. Gen.},\ }\textbf {\bibinfo {volume} {35}},\ \bibinfo {pages} {L481}
  (\bibinfo {year} {2002})}\BibitemShut {NoStop}%
\end{thebibliography}%

\end{document}